\documentclass[12pt]{article}
\usepackage{amsthm, bm, graphicx, hyperref, mathrsfs, bbm}
\usepackage{slashed}
\usepackage{amsfonts}
\usepackage{amsmath}
\usepackage{amssymb}
\usepackage{dsfont}
\usepackage{graphicx}
\usepackage{color}
\usepackage[all, knot]{xy}
\usepackage{tikz}
\usepackage{braket}
\usepackage{epstopdf}
\usepackage[footnotesize]{caption}
\usepackage{amsthm}
\usepackage{enumitem}
\usepackage{mathrsfs}
\usepackage{mathtools}     
\usepackage{subeqnarray}         
\usepackage{cases}               
\usepackage{color}
\usepackage{subfigure}
\usepackage{cite}               
\usepackage{hyperref}            
\usepackage{multirow,makecell}   
\usepackage{textcomp}
\usepackage{wasysym}
\usepackage{url}
\usepackage[margin=3cm]{geometry}
\usepackage{geometry}
\usepackage{amsfonts}
\usepackage{graphicx}
\usepackage{float}
\usepackage{amsmath}
\usepackage{hyperref}
\usepackage{tensor}
\usepackage{subfigure}
\usepackage{cases}
\usepackage{appendix}
\usepackage{multirow}
\usepackage{color}
\allowdisplaybreaks[3]

\begin{document}
\begin{titlepage}

\vspace{0.5cm}

\begin{center}

{\Large\bfseries
Emergent gravitational action from non-local \(T\bar T\)-like deformations
\par}

\vspace{0.5cm}

{\large
Yun-Ze Li$^{a}$
\footnote{lyz21@mails.jlu.edu.cn},
Bo-Rui Li$^{b,c},$%
\footnote{320220902891@lzu.edu.cn},
Yu-Xiao Liu$^{b,c,}$%
\footnote{liuyx@lzu.edu.cn (Corresponding author)},
Song He$^{a,}$%
\footnote{hesong@nbu.edu.cn (Corresponding author)}
\par}

\vspace{1.3em}

\begin{minipage}{0.90\textwidth}
\centering
\small\itshape
\setlength{\baselineskip}{1.18em}
\setlength{\parskip}{0.45em}

\noindent
$^{a}$Institute of Fundamental Physics and Quantum Technology, \& School of Physical Science and Technology,
Ningbo University, Ningbo, Zhejiang 315211, China
\par

\noindent
$^{b}$Key Laboratory of Quantum Theory and Applications of MoE,
Lanzhou Center for Theoretical Physics,
Key Laboratory of Theoretical Physics of Gansu Province,
Gansu Provincial Research Center for Basic Disciplines of Quantum Physics,
Lanzhou University, Lanzhou 730000, China
\par

\noindent
$^{c}$Institute of Theoretical Physics \& Research Center of Gravitation,
School of Physical Science and Technology,
Lanzhou University, Lanzhou 730000, China

\par

\end{minipage}

\end{center}

\vspace{0.9em}

\begin{center}
\begin{minipage}{0.94\textwidth}

\begin{center}
{\bfseries Abstract}
\end{center}

\vspace{-0.25em}

\small
\setlength{\baselineskip}{1.12em}
\setlength{\parindent}{0pt}
\setlength{\parskip}{0pt}

We study the gravitational effective action induced, at first order in the deformation parameter, by non-local \(T\bar T\)-like deformations of quantum field theories. Using a heat-kernel formulation, we extract the local geometric terms generated by stress-tensor two-point functions, and apply the construction to free fermions, massive Maxwell theory, and second-order Yang-Mills theory. While the resulting coefficients are generally model dependent, conformal field theories contain a universal sector fixed by the central charge \(C_T\). After regularization and renormalization, this sector yields finite, scheme-independent contributions organized into a finite set of curvature invariants. We further analyze trace-trace deformations, for which the relevant contact terms are determined by the Weyl anomaly, and derive the corresponding finite gravitational action for a general class of minimal non-local kernels. These results provide a quantum effective-action realization of induced gravity in which universal conformal data determine calculable contributions to the emergent geometric response.

\end{minipage}
\end{center}

\end{titlepage}
\newpage
\tableofcontents
\section{Introduction}
Understanding whether spacetime geometry and gravitational dynamics can be encoded in, or generated by, quantum field theoretic degrees of freedom is a longstanding question in theoretical physics. Holographic dualities, most notably AdS/CFT~\cite{Maldacena:1997re, Gubser:1998bc, Witten:1998qj}, provide a controlled realization of this idea by reconstructing bulk spacetime from boundary QFT data, although this framework relies on special large-$N$ limits, fixed asymptotic boundary conditions, and an additional higher-dimensional description. A complementary, non-holographic route is provided by Sakharov's induced gravity~\cite{Sakharov:1967nyk, Visser:2002ew, Adler:1982ri}, in which the gravitational action is not postulated as a fundamental input but arises as part of the quantum effective action obtained by integrating out matter fields on a curved background. In this setting, local geometric terms, including the cosmological constant, the Einstein-Hilbert term, and higher-curvature invariants, are generated by the response of quantum fields to the background metric. Related developments, such as the Adler-Zee formula and symmetry-breaking realizations of induced gravity~\cite{Adler:1980bx, Zee:1980sj}, further suggest that gravitational couplings can be encoded in stress-tensor correlation functions. At the same time, general constraints such as the Weinberg-Witten theorem~\cite{Weinberg:1980kq} indicate that obtaining a massless spin-two degree of freedom from an ordinary local Lorentz-covariant QFT is highly constrained. The focus of the present work is therefore not to construct a composite propagating graviton within a strictly local QFT, but rather to study the induced gravitational action, or equivalently, the geometric response encoded in quantum stress-tensor correlators. These considerations motivate settings in which the quantum origin of geometric terms can be analyzed explicitly, while relaxing locality assumptions in a controlled way.\par Stress-tensor deformations provide a natural framework for exploring this possibility. The stress tensor is the operator conjugate to the background metric, and its correlation functions therefore contain the quantum response data from which induced geometric terms may be extracted. In two dimensions, the solvable $T\bar T$ deformation~\cite{Zamolodchikov:2004ce,Smirnov:2016lqw,Cavaglia:2016oda} admits several geometric interpretations, including dynamical coordinate transformations, random geometry, and couplings to topological or massive gravity~\cite{Dubovsky:2017cnj,Dubovsky:2018bmo,Cardy:2018sdv,Tolley:2019nmm}. Higher-dimensional and generalized $T\bar T$-like deformations, including root-$T\bar T$ deformations, have also been investigated from various perspectives~\cite{Taylor:2018xcy, Bonelli:2018kik,Conti:2022egv,Babaei-Aghbolagh:2022uij,Ferko:2022cix,Morone:2024ffm,Babaei-Aghbolagh:2024hti}. Many existing geometric formulations of stress-tensor flows are classical or kinematical in character, relating the deformed theory to an undeformed theory on a modified metric, to auxiliary metric variables, or to a gravitational reframing of the dynamics. The viewpoint adopted here is closer in spirit to induced gravity: the stress tensor of the seed theory itself is used as the probe of geometry, and the induced gravitational action is derived from its quantum correlation functions. Non-local $T\bar T$-like deformations, obtained by inserting a differential or genuinely non-local kernel between two stress tensors, provide a concrete realization of this idea, with the kernel fixing the tensorial and derivative structure of the deformation and the stress-tensor correlators determining the resulting effective gravitational action.\par In our previous works~\cite{Li:2025lpa, Xie:2026kek}, we developed a semiclassical framework relating non-local $T\bar T$-like stress-tensor deformations to emergent gravitational dynamics, showing that such deformations admit an interpretation in terms of an effective gravitational saddle and that, in four-dimensional scalar examples, contact terms in the stress-tensor two-point function generate local geometric contributions to the induced action. The present paper focuses instead on the quantum effective-action aspect of this construction. Here the standard Sakharov contribution is understood as part of the undeformed seed-theory effective action at $\lambda=0$, whereas the quantity studied below is the $O(\lambda)$ change in its geometric sector generated by the prescribed non-local stress-tensor deformation. Our aim is to identify contributions to the emergent gravitational action that are universal, in the sense of being insensitive to both the microscopic realization of the seed theory and the choice of regularization scheme. We first construct a general framework applicable to free field theories and illustrate it in representative examples, including free fermions, massive Maxwell theory, and second-order Yang-Mills theory. In these free-field computations, the induced local geometric terms are extracted from the contact part of the stress-tensor two-point function, providing a unified computational setting while leaving the resulting structures model-dependent at the level of generic free theories. The main theory-independent results arise for conformal field theories in general dimensions, where conformal symmetry fixes the non-contact part of the stress-tensor two-point function up to the central charge $C_T$~\cite{OSBORN1994311}, allowing us to extract universal contributions to the induced gravitational action. By combining cutoff and dimensional regularization with a systematic renormalization procedure, we isolate finite, scheme-independent terms organized into a finite set of geometric invariants, rather than an infinite derivative expansion. A further universal sector arises from trace-trace deformations governed by the Weyl anomaly. In contrast to earlier treatments where the non-local kernel was fixed by the explicit form of the stress-tensor correlation function, we allow for a more general choice corresponding to arbitrary minimal operators and show that the anomaly determines additional finite contributions to the effective action. Together, these results sharpen the induced-gravity interpretation of non-local stress-tensor deformations by isolating renormalized, universal contributions to the effective gravitational action in general conformal field theories.\par
The rest of this paper is organized as follows. In section~\ref{sec:free}, we develop the general framework for non-local stress-tensor deformations of free field theories, discuss two regularization schemes, and apply the formalism to free fermions, massive Maxwell theory, and second-order Yang-Mills theory. In section~\ref{sec:cft}, we turn to conformal field theories in general dimensions, review the universal structure of the stress-tensor two-point function, analyze the deformation in dimensional regularization, and extract the scheme-independent part of the induced gravitational action. In section~\ref{sec:anomaly}, we study trace-trace deformations controlled by the Weyl anomaly, derive the corresponding trace-trace correlator, and determine the resulting finite gravitational action for general minimal kernels. We conclude in section~\ref{sec:conclusion} with a summary of the results and some open directions.
\section{Non-local stress tensor deformations of free field theories}\label{sec:free}
In this section, we aim to extract the effective gravitational action from the partition function of a free field theory under a non-local $T\bar T$-like deformation. We begin by presenting a general method for computing the first-order correction to the partition function, and explain how to recast it as a combination of local geometric invariants. We then explicitly evaluate the effective gravitational action in several concrete free-field theory models.
\subsection{Basic framework and general formulae}\label{Basic framework and general formulae}
Consider a $d$-dimensional manifold $\mathcal{M}$ without boundary, equipped with a generic metric $g$, on which a free field theory lives. The free field theory action is denoted by $S^{(0)}[g,\psi]$, where $\psi$ collectively represents the matter fields. The theory is deformed by a quadratic non-local stress tensor deformation; the resulting action is denoted by $S^{(\lambda)}[g,\psi]$ and satisfies the following flow equation,
\begin{align}
    \partial_{\lambda}S^{(\lambda)}[g,\psi]=\int_{\mathcal{M}\times\mathcal{M}} \text{d}\mu(x)\text{d}\mu(y)\,T_{\mu\nu}(x)H^{\mu\nu,\rho\sigma}(x,y)T_{\rho\sigma}(y),
\end{align}
where $\text{d}\mu(x)=\text{d}^dx\sqrt{g(x)}$ denotes the volume element. The non-local kernel $H^{\mu\nu,\rho\sigma}(x,y)$ can be expressed via the Green's function for a certain $M$-th order differential operator $F_{\mu\nu\alpha\beta}(x;\nabla)$,
\begin{align}\label{Green's function differential equation}
    F_{\mu\nu\alpha\beta}(x;\nabla)G^{\alpha\beta,\rho\sigma}(x,y)=\tilde\delta(x-y)\delta^{\rho}_{\mu}\delta^{\sigma}_{\nu},
\end{align}
with
\begin{align}\label{non-local kernel in deformation}
    H^{\mu\nu,\rho\sigma}(x,y)=I^{\mu\nu}_{\mu'\nu'}(x)G^{\mu'\nu',\rho'\sigma'}(x,y)J^{\rho\sigma}_{\rho'\sigma'}(y),
\end{align}
where $I^{\mu\nu}_{\mu'\nu'}$ and $J^{\rho\sigma}_{\rho'\sigma'}$ are projection maps which are constructed by some local geometric covariants. The operator $F_{\mu\nu\alpha\beta}(x;\nabla)$ only depends on the background metric and does not contain any information from the field theory. By definition, the flow equation of the deformed one-loop effective action takes the form
\begin{align}\label{flow equation for effective action}
    \partial_{\lambda}\mathcal{W}^{(\lambda)}[g]=\int \text{d}\mu(x)\text{d}\mu(y)\,H^{\mu\nu,\rho\sigma}(x,y)\langle{T_{\mu\nu}(x)T_{\rho\sigma}(y)}\rangle^{(\lambda)}.
\end{align}
This paper is primarily concerned with the first-order correction to the effective action,
\begin{align}\label{1-st effective action}
    \mathcal{W}^{(1)}[g]=\lambda\int \text{d}\mu(x)\text{d}\mu(y)\,H^{\mu\nu,\rho\sigma}(x,y)\langle{T_{\mu\nu}(x)T_{\rho\sigma}(y)}\rangle^{(0)},
\end{align}
which is governed by the explicit form of the non-local kernel and the two-point function of the stress tensor in the seed theory.\par
The local terms of the effective gravitational action are determined by the contact terms of the connected two-point function of the stress tensor. To extract these terms, one takes functional derivatives of the effective action of the seed theory with respect to the metric,
\begin{align}\label{definition of stress tensor 2-pt}
    \langle{T_{\mu\nu}(x)T_{\rho\sigma}(y)}\rangle^{(0)}=\frac{4}{\sqrt{g(x)}\sqrt{g(y)}}\Big(\frac{\delta^2\mathcal{W}^{(0)}}{\delta g^{\mu\nu}(x)\delta g^{\rho\sigma}(y)}+\frac{\delta\mathcal{W}^{(0)}}{\delta g^{\mu\nu}(x)}\frac{\delta\mathcal{W}^{(0)}}{\delta g^{\rho\sigma}(y)}\Big).
\end{align}
When the seed theory is chosen to be a free field theory, the one-loop effective action takes the form of the functional determinant of a certain $N$-th order (minimal) differential operator,
\begin{align}\label{1-loop effective action 1}
    \mathcal{W}^{(0)}[g]=\frac{1}{2}\ln\det \hat D,
\end{align}
where the operator $\hat D$, or equivalently $D^{A}_{B}$, is defined on the corresponding bundle of the matter field $\psi^{A}$. When $N$ is even, and $\hat D$ is minimal, the highest-order operator can be expressed via the Laplacian \cite{Barvinsky:1985an, Barvinsky:2021ijq},
\begin{align}
    \hat D=(-\Box)^{N/2}\hat{\mathbf{1}}+\hat P(\nabla).
\end{align}
$\hat P(\nabla)$ is composed of all lower-order operators, which can be formally written as
\begin{align}
    \hat P(\nabla)=\hat A_0+\hat A_1^{\mu_1}\nabla_{\mu_1}+\hat A_2^{\mu_1\mu_2}\nabla_{\mu_1}\nabla_{\mu_2}+\cdots+\hat A^{\mu_1\cdots\mu_{N-1}}_{N-1}\nabla_{\mu_1}\cdots\nabla_{\mu_{N-1}}.
\end{align}
The coefficients ${\hat A_p^{\mu_1\cdots\mu_p}}$ contain the information of the seed field theory. For instance, the coefficient $\hat A_0$ typically corresponds to the mass term of the field theory. The physical dimensions of these coefficients are
\begin{align}
    \dim \hat A_p^{\mu_1\cdots\mu_p}=N-p.
\end{align}
By employing the Mellin transformation, the one-loop effective action (\ref{1-loop effective action 1}) can be written as
\begin{align}\label{effective action of seed theory}
    \mathcal{W}^{(0)}[g]=-\frac{1}{2}\int_{0}^{\infty}\frac{\text{d}\tau}{\tau}\,\text{tr}(e^{-\tau\hat D}).
\end{align}
The trace $\text{tr}(e^{-\tau\hat D})$, which is also known as the trace of the heat kernel of $\hat D$, can be equivalently expressed as the integral 
\begin{align}
    \text{tr}(e^{-\tau\hat D})\equiv \int \text{d}\mu(x)\,K(\tau;x,x;\hat D).
\end{align}
In analogy with the Schwinger–DeWitt method \cite{Schwinger:1951nm,dewitt1965dynamical} for the heat kernel of a second-order differential operator, the diagonal components $K(\tau;x,x;\hat D)$ for the general-order minimal operator admits an asymptotic expansion in fractional powers of $\tau$ \cite{Barvinsky:2021ijq},
\begin{align}\label{expansion of seed theory heat kernel}
    K(\tau;x,x;\hat D)=\text{tr}_{V}\Big(\sum_{n=0}^{\infty}\tau^{\frac{2n-d}{N}}\hat E_{2n}(x)\Big).
\end{align}
The generalized Seeley-DeWitt coefficients $\{\hat E_{2n}\}$ are local functions of geometric invariants at $x$. Plugging (\ref{expansion of seed theory heat kernel}) into (\ref{effective action of seed theory}), and employing the proper-time regularization, we have
\begin{align}\label{UV divergent part of effective action}
    \mathcal{W}_{\Lambda}^{(0)}[g]&=-\frac{1}{2}\int \text{d}\mu(x)\,\text{tr}_{V}\Big(\sum_{n=0}^{\infty}\int_{\Lambda^{-N}}^{\infty}\text{d}\tau\,\tau^{\frac{2n-d}{N}-1}\hat E_{2n}(x)\Big)\notag\\
    &=-\frac{1}{2}\int \text{d}\mu(x)\,\text{tr}_{V}\Big(\sum_{n=0}^{d/2-1}\frac{N\Lambda^{d-2n}}{d-2n}\hat E_{2n}+N\ln(\Lambda/\mu_0)\hat E_{d}+O(\Lambda^0)\Big).
\end{align}
Here we assume that the spacetime dimension $d$ is an even number, and $\mu_0$ is a reference scale with physical dimension $\dim\mu_0=1$. According to the definition of stress tensor two-point function (\ref{definition of stress tensor 2-pt}), the second-order functional derivative of the UV-divergent part in (\ref{UV divergent part of effective action}) contributes to the contact terms in the correlation function, which can be formally written as
\begin{align}\label{general seed theory stress tensor two-point function}
     \langle{T_{\mu\nu}(x)T_{\rho\sigma}(y)}\rangle^{(0)}_{\text{cont.}}&=-\frac{2N\ln(\Lambda/\mu_0)}{\sqrt{g(x)g(y)}}\frac{\delta^2}{\delta g^{\mu\nu}(x)\delta g^{\rho\sigma}(y)}\int \text{d}\mu(z)\,\text{tr}_{V}\hat E_{d}(z)+\cdots\notag\\
     &=-2N\ln(\Lambda/\mu_0)E_{\mu\nu\rho\sigma}(x;\nabla)\tilde\delta(x-y)+\cdots.
\end{align}
Here, we have omitted the contact terms arising from the power-divergent parts, which are eliminated in the minimal subtraction renormalization scheme. The ellipsis in (\ref{general seed theory stress tensor two-point function}) contains the finite part of the two-point function, which includes contributions from non-contact terms that encode the long-distance information of the field theory. When combined with the non-local kernel of the deformation, these non-contact terms ultimately give rise to lower-order logarithmic divergences, which, after renormalization, yield a scheme-independent effective gravitational action. The computation of these contributions is carried out in the next section, where a conformal field theory is adopted as the seed theory. The coefficient $\hat E_{d}$ has physical dimension $\dim \hat E_{d}=d$, and can be decomposed into a linear combination of all possible terms of physical dimension $d$ constructed from products of the elements $\{\hat A_0,\hat A_{1}^{\mu_1},\cdots,\hat A^{\mu_1\cdots\mu_{N-1}}_{N-1},\hat {\mathbf{1}},g_{\mu\nu},g^{\mu\nu},\nabla_{\mu}, R^{\mu}_{\nu\rho\sigma},\hat{\mathcal{R}}_{\mu\nu}\}$, where $\hat{\mathcal{R}}_{\mu\nu}$ denotes the bundle curvature. $E_{\mu\nu\rho\sigma}(x;\nabla)$ is a certain differential operator built from these geometric invariants. Plugging (\ref{general seed theory stress tensor two-point function}) into (\ref{1-st effective action}), the first-order deformed effective action can be rewritten as
\begin{align}\label{1-st effective action 2}
    \mathcal{W}_{\Lambda}^{(1)}[g]=-2N\lambda\ln(\Lambda/\mu_0)\int \text{d}\mu(y)\lim_{x\to y}E^{\dagger}_{\mu\nu\rho\sigma}(x,\nabla)H^{\mu\nu,\rho\sigma}(x,y)+O(1).
\end{align}\par
The Green's function in (\ref{Green's function differential equation}) can be represented as a Mellin transform of the off-diagonal elements of the heat kernel for the operator $F_{\mu\nu\rho\sigma}(x;\nabla)$,
\begin{align}\label{zeta regularization for Green's function}
    G^{\mu\nu,\rho\sigma}(x,y)=\int_{0}^{\infty} \text{d}\tau\,K^{\mu\nu,\rho\sigma}(\tau;x,y;F),
\end{align}
where $K^{\mu\nu,\rho\sigma}(\tau;x,y;F)$ satisfies the differential equation
\begin{align}
    F_{\mu\nu\alpha\beta}(x;\nabla)K^{\alpha\beta,\rho\sigma}(\tau;x,y;F)=-\partial_{\tau}K_{\mu\nu}{}^{\rho\sigma}(\tau;x,y;F).
\end{align}
With $F_{\mu\nu\alpha\beta}(x;\nabla)$ assumed to be an $M$-th order minimal operator, the off-diagonal elements of its heat kernel can be expanded in fractional powers of $\tau$ as \cite{Barvinsky:2021ijq},
\begin{align}\label{series expansion for heat kernel of F}
    K^{\mu\nu,\rho\sigma}(\tau;x,y;F)&=\frac{\Delta^{-1}(x,y)}{(4\pi )^{d/2}}\sum_{p=-\infty}^{\infty}\tau^{\frac{2p-d}{M}}\notag\\
    &\quad\times\sum_{q\geq N_p}\mathcal{E}_{\frac{M}{2},\frac{d+Mq-2p}{2}}\Big(-\frac{\sigma}{2\tau^{2/M}}\Big)b^{\mu\nu,\rho\sigma}_{p,q}(x,y),
\end{align}
where $\sigma$ is the Synge world function and $\Delta$ is the Van Vleck–Morette determinant. $\mathcal{E}_{\nu,\alpha}$ is called the generalized exponential function defined by power series
\begin{align}
    \mathcal{E}_{\nu,\alpha}(z)=\frac{1}{\nu}\sum_{l=0}^{\infty}\frac{\Gamma(\frac{\alpha+l}{\nu})}{\Gamma(\alpha+l)}\frac{z^l}{l!}.
\end{align}
The constant $N_p=2p/M$ when $p\geq 0$, and $N_p=2|p|/(M-1)$ when $p<0$. Now we need to substitute (\ref{zeta regularization for Green's function}) and (\ref{series expansion for heat kernel of F}) back into the coincidence limit in the integral (\ref{1-st effective action 2}). When performing the integration over $\tau$ in (\ref{zeta regularization for Green's function}), the integral always converges at the $\tau\to0^+$ side owing to the presence of the generalized exponential function in the heat kernel expansion (\ref{series expansion for heat kernel of F}). However, upon taking the coincidence limit $x\to y$, the Synge world function $\sigma(x,y)$ tends to zero, which leads to the degeneracy of the generalized exponential function and hence to the divergence of the integral as $\tau \to 0^+$. We may therefore introduce a UV cutoff $\tau=\Lambda^{-M}$ as the lower limit of integration. With this cutoff in place, the integration over the truncated region and the coincidence limit can be interchanged,
\begin{align}\label{Green's function interchange}
&\quad\lim_{x\to y}E^{\dagger}_{\mu\nu\rho\sigma}(x,\nabla)H_{\Lambda}^{\mu\nu,\rho\sigma}(x,y)\notag\\
&=\int_{\Lambda^{-M}}^{\infty}\text{d}\tau\,\lim_{x\to y}E^{\dagger}_{\mu\nu\rho\sigma}(x,\nabla)\Big(I^{\mu\nu}_{\mu'\nu'}(x)K^{\mu'\nu',\rho'\sigma'}(\tau;x,y;F)J^{\rho\sigma}_{\rho'\sigma'}(y)\Big).
\end{align}
By acting with the covariant differential operator $E^{\dagger}_{\mu\nu\rho\sigma}(x,\nabla)$ on the heat kernel expansion (\ref{series expansion for heat kernel of F}) and then taking the coincidence limit $x\to y$, the integrand on the right-hand side of (\ref{Green's function interchange}) can be expressed as a fractional power series in $\tau$,
\begin{align}\label{Green's function coincidence limit}
    \lim_{x\to y}E^{\dagger}_{\mu\nu\rho\sigma}(x,\nabla)\Big(I^{\mu\nu}_{\mu'\nu'}(x)K^{\mu'\nu',\rho'\sigma'}(\tau;x,y;F)J^{\rho\sigma}_{\rho'\sigma'}(y)\Big)&=\sum_{m=-\infty}^{\infty}\tau^{\frac{2m-d}{M}}B_{2m}(y),
\end{align}
where the coefficients $\{B_{2m}\}$ are built from geometric invariants. Finally, by plugging (\ref{Green's function interchange}) and (\ref{Green's function coincidence limit}) into (\ref{1-st effective action 2}), we obtain the leading logarithmic divergence in the first-order deformed effective action,
\begin{align}\label{UV cutoff regularization leeading log div}
    \mathcal{W}_{\Lambda}^{(1)}[g]&=-2MN\ln^2(\Lambda/\mu_0)\int \text{d}\mu(y)B_{d-M}(y)+O(\ln(\Lambda/\mu_0)).
\end{align}
Here we have omitted the power-divergent parts. After renormalization, a finite, scheme-independent effective gravitational action is obtained from these logarithmic divergences.
\subsection{Another regularization scheme}
In this subsection, the logarithmic divergences in the first-order deformed effective action are reproduced by means of dimensional regularization, which is employed as an alternative regularization scheme. The discussion in this subsection begins with the effective action (\ref{effective action of seed theory}) of the free field theory. Instead of the UV cutoff regularization introduced earlier, we adopt dimensional regularization, $d \mapsto D = d - \varepsilon$, which is more convenient. To exclude the contribution from IR divergence, an upper limit $\tau=L^N$ is introduced to divide the UV and IR regions. By substituting the heat kernel expansion (\ref{expansion of seed theory heat kernel}) into (\ref{effective action of seed theory}), one obtains
\begin{align}
       \mathcal{W}_{\text{reg}}^{(0)}[g]&=-\frac{1}{2}\int \text{d}\mu(x)\,\text{tr}_{V}\Big(\sum_{n=0}^{\infty}\int_{0}^{L^{N}}\text{d}\tau\,\mu^{\varepsilon}\tau^{\frac{2n-D}{N}-1}\hat E_{2n}(x)\Big)\notag\\
       &=-\frac{1}{2}\int \text{d}\mu(x)\,\text{tr}_{V}\Big(\mu^{\varepsilon}\sum_{n=0}^{\infty}\frac{NL^{2n-D}}{2n-D}\hat E_{2n}(x)\Big),
\end{align}
where the prefactor $\mu^{\varepsilon}$ is introduced to balance the physical dimension. The regularized stress tensor two-point functions are obtained by taking functional derivatives of $\mathcal{W}_{\text{reg}}^{(0)}$ with respect to the metric,
\begin{align}
    \langle{T_{\mu\nu}(x)T_{\rho\sigma}(y)}\rangle^{(0)}_{\text{reg}}&=-\frac{2}{\sqrt{g(x)g(y)}}\frac{\delta^2}{\delta g^{\mu\nu}(x)\delta g^{\rho\sigma}(y)}\int \text{d}\mu(z)\,\text{tr}_{V}\Big(\mu^{\varepsilon}\sum_{n=0}^{\infty}\frac{NL^{2n-D}}{2n-D}\hat E_{2n}(z)\Big)\notag\\
     &=-\mu^{\varepsilon}\sum_{n=0}^{\infty}\frac{NL^{2n-D}}{2n-D}E^{(2n)}_{\mu\nu\rho\sigma}(x;\nabla)\tilde\delta(x-y).
\end{align}
Plugging it back into the first-order deformed effective action (\ref{1-st effective action}) and using the Mellin transformation (\ref{zeta regularization for Green's function}), we have
\begin{align}
        \mathcal{W}^{(1)}_{\text{reg}}[g]=-\lambda\mu^{\varepsilon}\sum_{n=0}^{\infty}\frac{NL^{2n-D}}{2n-D}\int \text{d}\mu(y)\,\lim_{x\to y}\int_{0}^{R^M}\text{d}\tau\,E^{(2n)\dagger}_{\mu\nu\rho\sigma}(x;\nabla)\tilde K^{\mu\nu,\rho\sigma}(\tau;x,y;F),
\end{align}
where $\tilde K^{\mu\nu,\rho\sigma}(\tau;x,y;F)=I^{\mu\nu}_{\mu'\nu'}(x)K^{\mu'\nu',\rho'\sigma'}(\tau;x,y;F)J^{\rho\sigma}_{\rho'\sigma'}(y)$ and the upper limit $R^M$ is an arbitrary scale. Next, we need to interchange the order of the $\tau$-integration and the limit $x\to y$. However, according to the heat kernel expansion (\ref{series expansion for heat kernel of F}), the generalized exponential function $\mathcal{E}\bigl(-\frac{\sigma}{2\tau^{2/M}}\bigr)$ degenerates in the coincidence limit, and the negative integer powers of $\tau$ appearing in the expansion then give rise to divergences in the integral at $\tau=0$. To resolve this, we introduce dimensional regularization $d\mapsto D = d-\varepsilon$, which shifts the powers of $\tau$ in the expansion, so that the integrals with negative powers are rendered finite via analytic continuation,
\begin{align}
    \lim_{x\to y}\int_{0}^{R^M}\text{d}\tau\,E^{(2n)\dagger}_{\mu\nu\rho\sigma}(x;\nabla)\tilde K_{\varepsilon}^{\mu\nu,\rho\sigma}(\tau;x,y;F)&=\mu^{\varepsilon}\int_{0}^{R^{M}}\text{d}\tau\,\sum_{m=-\infty}^{\infty}\tau^{\frac{2m-D}{M}}B_{2m}^{(2n)}(y)\notag\\
    &=\mu^{\varepsilon}\sum_{m=-\infty}^{\infty}\frac{MR^{2m+M-D}}{2m+M-D}B_{2m}^{(2n)}(y).
\end{align}
Plugging it back into the first-order deformed effective action, one obtains
\begin{align}
    \mathcal{W}^{(1)}_{\text{reg}}[g]&=-\lambda\sum_{n=0}^{\infty}\sum_{m=-\infty}^{\infty}\mu^{2\varepsilon}\frac{MNL^{2n-D}R^{2m+M-D}}{(2n-D)(2m+M-D)}\int \text{d}\mu(y)B_{2m}^{(2n)}(y),\notag\\
    &=-\lambda MN\bigg\{\Big(\frac{1}{\varepsilon^2}+\frac{\ln(\mu^2LR)}{\varepsilon}+\frac{1}{2}\ln^2(\mu^2LR)\Big)\int \text{d}\mu(y)\,B_{d-M}^{(d)}(y)\notag\\
    &\quad+\sum_{n\neq d/2}\frac{L^{2n-d}}{2n-d}\Big(\frac{1}{\varepsilon}+\ln(\mu^2LR)-\frac{1}{2n-d}\Big)\int \text{d}\mu(y)\,B_{d-M}^{(2n)}(y)\notag\\
    &\quad+\sum_{m\neq (d-M)/2}\frac{R^{2m+M-d}}{2m+M-d}\Big(\frac{1}{\varepsilon}+\ln(\mu^2LR)-\frac{1}{2m+M-d}\Big)\int \text{d}\mu(y)\,B_{2m}^{(d)}(y)\notag\\
    &\quad+\sum_{\substack{n\neq d/2\\m\neq (d-M)/2}}\frac{L^{2n-d}R^{2m+M-d}}{(2n-d)(2m+M-d)}\int \text{d}\mu(y)\,B_{2m}^{(2n)}(y)\bigg\}+O(\varepsilon).
\end{align}
The renormalized effective action is obtained via minimal subtraction, with all negative powers of $\varepsilon$ subtracted. Among the remaining finite terms, some are proportional to powers of the unphysical parameters $(L, R)$ and are therefore scheme-dependent. The scheme-independent finite part, in contrast, arises from the term in $W^{(1)}_{\text{reg}}$ that is proportional to $\ln^2\mu$,
\begin{align}
    W^{(1)}_{\text{ren,SI}}[g]=-2MN\ln^2(\mu/\mu_0)\int \text{d}\mu(y)\,B_{d-M}^{(d)}(y),
\end{align}
which is consistent with the leading logarithmic divergence (\ref{UV cutoff regularization leeading log div}) obtained in the UV cutoff regularization.
\subsection{Examples}
In the following examples, we specialize to the non-local kernel derived from the linearized Einstein-gravity construction of \cite{Li:2025lpa}, which means we take the projection maps in \eqref{non-local kernel in deformation} in the following form:
\begin{equation}
\begin{aligned}
    I^{\mu\nu}_{\mu'\nu'}(x)&= \delta^{\mu}_{\mu'}\delta^{\nu}_{\nu'}, \\
J^{\rho\sigma}_{\rho'\sigma'}(y) &= \delta^{\rho}_{\rho'}\delta^{\sigma}_{\sigma'} - \frac{1}{2}\, g_{\rho'\sigma'}(y) g^{\rho\sigma}(y),
\end{aligned}
\end{equation}
and the differential operator defining the Green's function as
\begin{equation}
    F_{\mu\nu\alpha\beta}(x;\nabla) = -g_{\mu\rho}g_{\nu\sigma}\,\Box_x - R_{\mu\rho\nu\sigma} - R_{\nu\rho\mu\sigma}.
\end{equation}
We evaluate the first-order effective action for a free fermion, a massive Maxwell field, and second-order Yang-Mills theory in four dimensions. These examples illustrate how the contact part of the stress-tensor two-point function induces local gravitational terms whose form depends on the microscopic structure of the seed theory.

\subsubsection{Free fermion in four dimensions}
\label{sec:dirac}

We first consider a massive Dirac field on a four-dimensional Euclidean manifold.  Besides extending the analysis to fermionic matter, this example provides a useful test of how the intrinsic mass scale and the spin-connection curvature enter the induced gravitational action.  The Euclidean action is
\begin{equation}
\label{eq:dirac-action}
S_{F}[\psi,\bar\psi,e]
=
\int  \text{d}^4x\,\sqrt{g}\,
\bar\psi\left(\slashed{\nabla}+m\right)\psi,
\quad
\slashed{\nabla}=\gamma^{a}\nabla_{a},
\quad
\nabla_{a}=e_{a}{}^{\mu}\left(\partial_{\mu}+\Omega_{\mu}\right),
\end{equation}
where \(e_{a}{}^{\mu}\) is the vielbein and \(\Omega_{\mu}\) is the spin connection.  The Grassmann integral gives
\begin{equation}
\label{eq:dirac-determinant}
\mathcal{Z}
=
\det\!\left(\slashed{\nabla}+m\right),
\qquad
\mathcal{W}
=
-\ln \mathcal{Z}.
\end{equation}
For the parity-even local part of the effective action, the Dirac operator may be squared.  Using the Lichnerowicz formula~\cite{Lichnerowicz:1963,Lawson:1989},
\begin{equation}
\label{eq:lichnerowicz}
-\slashed{\nabla}^{\,2}
=
-\nabla^{2}+\frac{1}{4}R,
\end{equation}
we obtain
\begin{equation}
\label{eq:dirac-laplace-operator}
\mathcal{W}
=
-\frac{1}{2}\operatorname{Tr}\ln D_{F},
\qquad
D_{F}
=
-\nabla^{2}+\frac{1}{4}R+m^{2}.
\end{equation}

In the standard convention for a Laplace-type operator,
\(D=-(\nabla^{2}+E)\), the endomorphism and bundle curvature are
\begin{equation}
\label{eq:dirac-bundle-data}
E
=
-\left(m^{2}+\frac{1}{4}R\right)\hat{\mathbf 1},
\quad
\Omega_{\mu\nu}
=
\frac{1}{4}R_{\mu\nu ab}\gamma^{ab},
\quad
\operatorname{tr}
\left(\Omega_{\mu\nu}\Omega^{\mu\nu}\right)
=
-\frac{1}{2}R_{\mu\nu\rho\sigma}R^{\mu\nu\rho\sigma}.
\end{equation}
The standard heat-kernel coefficients for these data~\cite{Vassilevich:2003xt} give the regulated effective action
\begin{equation}
\label{eq:dirac-effective-action}
\begin{aligned}
\mathcal{W}_{\Lambda}
=&
\frac{1}{16\pi^{2}}
\int  \text{d}^4x\,\sqrt{g}\,
\Bigg[
\Lambda^{4}
-\frac{\Lambda^{2}}{6}\left(R+12m^{2}\right)
\\
&\quad+2\ln(\Lambda/m)
\left(
m^{4}
+\frac{1}{6}m^{2}R
+\frac{1}{144}R^{2}
-\frac{1}{90}R_{\mu\nu}R^{\mu\nu}
-\frac{7}{720}
R_{\mu\nu\rho\sigma}R^{\mu\nu\rho\sigma}
\right)
\Bigg]
+O(\Lambda^{0}),
\end{aligned}
\end{equation}
where total derivatives have been omitted on the boundaryless manifold considered here.  The quartic and quadratic divergences renormalize local operators of dimension zero and two, while the logarithmic term controls the scheme-independent contribution after its contraction with the non-local kernel.

Taking two metric variations of \eqref{eq:dirac-effective-action} gives the contact part of the stress-tensor two-point function.  Up to second order in derivatives, it may be written as
\begin{align}\label{eq:tt-contact}
\left\langle
T^{\mu\nu}(x)T^{\rho\sigma}(y)
\right\rangle
=&
\frac{1}{16\pi^{2}}
\Big\{
\left(
\Lambda^{4}
-2m^{2}\Lambda^{2}
+2m^{4}\ln(\Lambda/m)
\right)
\left(
g^{\mu\nu}g^{\rho\sigma}
-2g^{\mu(\rho}g^{\sigma)\nu}
\right)
\notag\\
&\quad-\frac{1}{3}
\left(
\Lambda^{2}
-2m^{2}\ln(\Lambda/m)
\right)
\Big[
2\left(
g^{\mu\nu}g^{\rho\sigma}
+g^{\mu(\rho}g^{\sigma)\nu}
\right)\Box
\notag\\
&\quad
-2g^{\mu\nu}\nabla^{(\rho}\nabla^{\sigma)}
-2g^{\rho\sigma}\nabla^{(\mu}\nabla^{\nu)}
-4R^{\mu\rho\nu\sigma}
\notag\\
&\quad
+2g^{\mu\nu}R^{\rho\sigma}
+2g^{\rho\sigma}R^{\mu\nu}
-2g^{\mu(\rho}R^{\sigma)\nu}
-2g^{\nu(\rho}R^{\sigma)\mu}
\Big]
\Big\}
\delta(x-y)
\notag\\
&\quad
+\ln(\Lambda/m)
F_{\rm Dirac}^{(4)\,\mu\nu\rho\sigma}(x,y)
+O(\Lambda^{0}).
\end{align}
All covariant derivatives in this expression act on \(\delta(x-y)\) with respect to \(x\). The tensor
\(F_{\rm Dirac}^{(4)\,\mu\nu\rho\sigma}\) collects the fourth-derivative contact terms generated by the curvature-squared sector of \eqref{eq:dirac-effective-action}.  At the derivative order retained below, only the explicitly displayed zero- and two-derivative terms are required.

We now insert \eqref{eq:tt-contact} into the general first-order deformation \eqref{1-st effective action}. Its regulated coincidence limits are obtained from the heat-kernel representation developed in subsection~\ref{Basic framework and general formulae}. By employing the UV-cutoff regularization, the regularized Green's function can be written as
\begin{align}\label{eq:g-exp}
G^{\mu\nu,\rho\sigma}_{\Lambda}(x,y)=\int_{\Lambda^{-2}}^{\infty}\text{d}\tau\,\frac{\Delta^{1/2}(x,y)}{16\pi^2\tau^2}e^{-\sigma(x,y)/2\tau}\sum_{n=0}^{\infty}a^{\mu\nu,\rho\sigma}_{n}(x,y)\tau^n.
\end{align}
Since the correlator in \eqref{eq:tt-contact} is supported at coincident points, the bilocal integral collapses to local coincidence limits of \(H_{\mu\nu\rho\sigma}\) and its covariant derivatives. The regulated coincidence limit reads 
\begin{equation}\label{massive vector green function} 
G_{\mu\nu\rho\sigma}^{\Lambda}(x,x) = \frac{1}{16\pi^{2}} \left[ \Lambda^{2}\mathbb{I}_{\mu\nu\rho\sigma} + 2\ln\Lambda \left( R_{\mu\rho\nu\sigma} +R_{\nu\rho\mu\sigma} +\frac{1}{6}R\,\mathbb{I}_{\mu\nu\rho\sigma} \right) +\cdots \right], 
\end{equation}
where $\mathbb{I}_{\mu\nu\rho\sigma} = g_{\mu(\rho}g_{\sigma)\nu}$ is the identity on symmetric rank-two tensors. The corresponding derivative coincidence limits follow from the same heat-kernel expansion and generate local curvature-squared and total-derivative terms.

Plugging (\ref{eq:tt-contact})(\ref{massive vector green function})(\ref{eq:g-exp}) into (\ref{1-st effective action}), we can rewrite the first-order deformed effective action as
\begin{align}\label{eq:local-h-reduction}
    \mathcal{W}^{(1)}_{\Lambda}&=\frac{\lambda m^2\ln^2(\Lambda/m)}{128\pi^4}\int \text{d}^4x\sqrt{g}\Big[-\frac{20m^2}{3}R+\frac{10}{9}R^2+\frac{112}{45}R_{\mu\nu}R^{\mu\nu}-8R_{\mu\nu\rho\sigma}R^{\mu\nu\rho\sigma}\notag\\
    &\quad+\frac{88}{45}\Box R+O(R^3,\nabla^6)\Big]+\text{power-law divergences}.
\end{align}
On a compact manifold without boundary, or under suitable boundary conditions, the total derivative $\Box R$ may be discarded. The power-law divergent pieces renormalize the corresponding local gravitational counterterms, whereas the logarithmically enhanced terms determine the finite, scheme-independent sector after renormalization:
\begin{align}
     \mathcal{W}^{(1)}_{\text{ren}}&=\frac{\lambda m^2\ln^2(\mu/m)}{128\pi^4}\int \text{d}^4x\sqrt{g}\Big[-\frac{20m^2}{3}R+\frac{10}{9}R^2+\frac{112}{45}R_{\mu\nu}R^{\mu\nu}-8R_{\mu\nu\rho\sigma}R^{\mu\nu\rho\sigma}\Big)\notag\\
     &\quad+O(R^3,\nabla^6)\Big],
\end{align}
where $\mu$ is the renormalization scale. In particular, the fermion mass supplies the intrinsic scale required for lower-derivative geometric terms, while the spinor bundle curvature modifies the coefficients of the curvature-squared sector relative to the bosonic examples.

\subsubsection{Proca field theory in four dimensions}

We next consider a massive Abelian vector field in four-dimensional Euclidean spacetime, described by the Proca action 
\begin{equation}\label{massive vector field}
S_{\mathrm{Proca}}
=
\int \text{d}^4x\,\sqrt{g}\,
\left(
\frac{1}{4}F_{\mu\nu}F^{\mu\nu}
+\frac{1}{2}m^{2}A_{\mu}A^{\mu}
\right),
\end{equation}
where
\begin{equation}
F_{\mu\nu}=\nabla_{\mu}A_{\nu}-\nabla_{\nu}A_{\mu}.
\end{equation}
After integration by parts, the action takes the quadratic form
\begin{equation}
S_{\mathrm{Proca}}
=
\frac{1}{2}
\int \text{d}^4x\,\sqrt{g}\,
A_{\mu}D^{\mu\nu}A_{\nu},
\quad
D^{\mu\nu}
=
-g^{\mu\nu}\nabla^{2}
+\nabla^{\mu}\nabla^{\nu}
+m^{2}g^{\mu\nu}
+R^{\mu\nu}.
\end{equation}
Unlike the Laplace-type operators encountered for minimally coupled scalar fields, $D^{\mu\nu}$ is non-minimal because of the term $\nabla^{\mu}\nabla^{\nu}$. This feature reflects the longitudinal polarization of the massive vector field, leading to a distinct set of curvature coefficients in the one-loop effective action.

Since the theory is Gaussian, its one-loop effective action is exact. The partition function is given by
\begin{equation}\label{MM partition function}
\mathcal{Z}
=
\int\mathcal{D}A_{\mu}\,e^{-S_{\mathrm{Proca}}}
=
\det{}^{-1/2}D^{\mu\nu}
=
\exp\left(
-\frac{1}{2}\ln\det D^{\mu\nu}
\right).
\end{equation}
Using a proper-time cutoff $\Lambda$, the regulated one-loop effective action takes the form 
\begin{align}
\mathcal{W}_{\Lambda} &= \frac{\ln(\Lambda/m)}{(4\pi)^{2}} \int \text{d}^4x\,\sqrt{g}\, \bigg[ \frac{3}{2}m^{4} +\frac{1}{2}m^{2}R -\frac{1}{8}R^{2} \notag\\
&\quad +\frac{29}{60}R_{\mu\nu}R^{\mu\nu} -\frac{1}{15}R_{\mu\nu\rho\sigma}R^{\mu\nu\rho\sigma} \bigg].
\end{align}
For the first-order deformation considered below, the terms proportional to $m^{4}$ and $m^{2}R$ determine the contact contribution up to second order in derivatives, whereas the curvature-squared terms contribute only at higher derivative order after taking two metric variations. The contact contribution to the stress-tensor two-point function is obtained by taking two metric variations of the logarithmic part of the effective action.  Up to second order in derivatives, one finds
\begin{align}
\left\langle
T^{\mu\nu}(x)T^{\rho\sigma}(y)
\right\rangle
={}&
\frac{\ln(\Lambda/m)}{(4\pi)^{2}}
\Big\{
3m^{4}
\left(
g^{\mu\nu}g^{\rho\sigma}
-2g^{\mu(\rho}g^{\sigma)\nu}
\right)
\notag\\
&\quad
+m^{2}\Big[
g^{\mu(\rho}g^{\sigma)\nu}\Box
-g^{\mu\nu}g^{\rho\sigma}\Box
+g^{\mu\nu}\nabla^{(\rho}\nabla^{\sigma)}
+g^{\rho\sigma}\nabla^{(\mu}\nabla^{\nu)}
\notag\\
&\quad
-2g^{\mu(\rho}\nabla^{\sigma)}\nabla^{\nu}
-2g^{\nu(\rho}\nabla^{\sigma)}\nabla^{\mu}
-2R^{\mu\rho\nu\sigma}
-2R^{\nu\rho\mu\sigma}
\notag\\
&\quad
+4g^{\mu\nu}R^{\rho\sigma}
+4g^{\rho\sigma}R^{\mu\nu}
-2g^{\mu(\rho}R^{\sigma)\nu}
-2g^{\nu(\rho}R^{\sigma)\mu}
\Big]
+O(\nabla^{4})
\Big\}
\delta(x-y)
\notag\\
&\quad
+\text{power-law divergences},
\end{align}
where all covariant derivatives act on $\delta(x-y)$ with respect to $x$. Substituting these results into the first-order deformation gives
\begin{align}
\mathcal{W}^{(1)}_{\Lambda} &=\frac{\lambda m^2\ln(\Lambda/m)}{64\pi^{4}} \int \text{d}^4x\,\sqrt{g}\, \Big[ \frac{15m^{2}\Lambda^{2}}{4} + \Big( \frac{m^{2}\ln(\Lambda/m)}{2} +\Lambda^2\Big)R \notag\\ 
&\quad+ \frac{\ln(\Lambda/m)}{16} \Big( 12R_{\mu\nu\rho\sigma}R^{\mu\nu\rho\sigma} -24R_{\mu\nu}R^{\mu\nu} +12R^{2} -\frac{19}{15}\Box R \Big) \notag\\
&\quad+O(R^{3},\nabla^{6}) \Big].
\end{align} 
The terms proportional to $\Lambda^{2}$ are local power-law divergences and can be absorbed into gravitational counterterms. The logarithmically enhanced part is proportional to $\ln^{2}(\Lambda/m)$ and determines the scheme-independent contribution considered below. 

After subtracting the power-law divergences and discarding total derivatives, the logarithmically enhanced contribution to the first-order gravitational action is
\begin{align}
\mathcal{W}^{(1)}_{\mathrm{ren}} &= \frac{\lambda m^{2}}{128\pi^{4}}\ln^2(\mu/m) \int \text{d}^4x\,\sqrt{g}\, \left[ 2m^{2}R + 3\left( R_{\mu\nu\rho\sigma}R^{\mu\nu\rho\sigma} -2R_{\mu\nu}R^{\mu\nu} +R^{2} \right) \right] \notag\\
&\quad+O(R^{3},\nabla^{6}).
\end{align}
The curvature-squared combination may be written as $E_{4}+2R_{\mu\nu}R^{\mu\nu}$, where $E_{4}$ is the four-dimensional Euler density. The deformation therefore induces both an Einstein-Hilbert term and curvature-squared corrections. Matching the coefficient of $R$ to the standard normalization $(16\pi G_{\mathrm{eff}})^{-1}$ gives 
\begin{align}
G_{\mathrm{eff}}=\frac{4\pi^{3}}{\lambda m^{4}\ln^{2}(\mu/m)}.
\end{align}
The coefficient of the Einstein-Hilbert term scales as $m^{4}$, showing that the induced Newton coupling is set by the intrinsic mass scale of the Proca seed theory.

It is worth emphasizing that the Maxwell field theory cannot be obtained by simply setting \(m=0\) in the Proca result, since its quantization requires gauge fixing and the inclusion of the corresponding Faddeev-Popov ghost determinant. The logarithmically divergent part of the Maxwell one-loop effective action starts at quadratic order in the curvature, so that its second metric variation is already of fourth derivative order. After contraction with the same non-local kernel and renormalization of the local divergent terms, the first-order correction therefore begins at \(O(R^3,\nabla^{6})\). In particular, no Einstein-Hilbert or curvature-squared contribution is generated at this order.

\subsubsection{Second-order Yang-Mills theory in four dimensions}

We finally consider non-Abelian Yang-Mills theory in four-dimensional Euclidean spacetime,
\begin{equation}\label{YM action}
S_{\mathrm{YM}}
=
\frac{1}{4}
\int \text{d}^4x\,\sqrt{g}\,
\gamma_{ab}F_{\mu\nu}^{a}F^{b\mu\nu},
\qquad
F_{\mu\nu}^{a}
=
\partial_{\mu}A_{\nu}^{a}
-\partial_{\nu}A_{\mu}^{a}
+C^{a}{}_{bc}A_{\mu}^{b}A_{\nu}^{c},
\end{equation}
where $C^{a}{}_{bc}$ are the structure constants and
$\gamma_{ab}=C^{f}{}_{ad}C^{d}{}_{bf}$ is the Killing metric of the gauge algebra.  For $G=\mathrm{SU}(N)$, we adopt the conventional normalization
$\gamma_{ab}=\delta_{ab}$ and denote the dimension of the gauge group by
$n=\delta^{a}_{a}=\dim G$.

Since the non-Abelian theory is interacting, its action is not quadratic in the gauge field.  The one-loop effective action is nevertheless determined by the quadratic fluctuations around a classical background.  We therefore introduce the background-field decomposition
\begin{equation}
A_{\mu}^{a}
=
\mathbb{A}_{\mu}^{a}+h_{\mu}^{a},
\qquad
\mathbb{F}_{\mu\nu}^{a}
=
\partial_{\mu}\mathbb{A}_{\nu}^{a}
-\partial_{\nu}\mathbb{A}_{\mu}^{a}
+C^{a}{}_{bc}\mathbb{A}_{\mu}^{b}\mathbb{A}_{\nu}^{c},
\end{equation}
together with the background-covariant derivative
\begin{equation}\label{total derivative}
\nabla_{\mu}h_{\nu}^{a}
=
\nabla_{\mu}^{(\mathrm{R})}h_{\nu}^{a}
+C^{a}{}_{bc}\mathbb{A}_{\mu}^{b}h_{\nu}^{c},
\end{equation}
where $\nabla_{\mu}^{(\mathrm{R})}$ acts on the spacetime index.  Expanding around a background satisfying the classical Yang-Mills equations, the linear term vanishes and
\begin{equation}\label{expand to second order}
S_{\mathrm{YM}}[\mathbb{A}+h]
=
S_{\mathrm{YM}}[\mathbb{A}]
+
\frac{1}{2}
\int \text{d}^4x\,\sqrt{g}\,
h_{\mu}^{a}O^{\mu\nu}{}_{ab}h_{\nu}^{b}
+O(h^{3}),
\end{equation}
where
\begin{equation}
O^{\mu\nu}{}_{ab}
=
-\gamma_{ab}g^{\mu\nu}\nabla^{2}
+\gamma_{ab}\nabla^{\mu}\nabla^{\nu}
+\gamma_{ab}R^{\mu\nu}
+2\gamma_{ac}C^{c}{}_{db}\mathbb{F}^{\text{d}\mu\nu}.
\end{equation}
Thus, the ``second-order'' theory considered here refers to the quadratic background-field sector relevant for the one-loop determinant; the interaction vertices contained in $O(h^{3})$ do not contribute at this order.

We impose the background-covariant Feynman gauge,
\begin{equation}\label{gauge fixing}
\begin{aligned}
S_{\mathrm{gf}}
&=
\frac{1}{2}
\int \text{d}^4x\,\sqrt{g}\,
\gamma_{ab}
\left(\nabla_{\mu}h^{a\mu}\right)
\left(\nabla_{\nu}h^{b\nu}\right),
\\
S_{\mathrm{gh}}
&=
\int \text{d}^4x\,\sqrt{g}\,
\bar c_{a}(-\nabla^{2})c^{a}.
\end{aligned}
\end{equation}
After integration by parts, the gauge-fixing term cancels the non-minimal contribution
$\nabla^{\mu}\nabla^{\nu}$ in $O^{\mu\nu}{}_{ab}$.  The quadratic gauge-field operator therefore reduces to the minimal Laplace-type operator
\begin{equation}
D^{\mu\nu}{}_{ab}
=
-\gamma_{ab}g^{\mu\nu}\nabla^{2}
+\gamma_{ab}R^{\mu\nu}
+2\gamma_{ac}C^{c}{}_{db}\mathbb{F}^{\text{d}\mu\nu}.
\end{equation}
The gauge and ghost fluctuations then give
\begin{equation}
\mathcal{Z}^{\text{1-loop}}
=
\det(-\nabla^{2})\,
\det{}^{-1/2}D,
\quad
\mathcal{W}
=
-\ln\det(-\nabla^{2})
+\frac{1}{2}\ln\det D.
\end{equation}

Using the proper-time regularization, the logarithmically divergent part of the one-loop effective action is given by~\cite{Barvinsky:1985an}
\begin{align}
\mathcal{W}_{\Lambda}
&=
\int_{\Lambda^{-2}}^{\infty}
\frac{\text{d}\tau}{\tau}
\left[
\operatorname{Tr}\!\left(e^{-\tau(-\nabla^{2})}\right)
-\frac{1}{2}\operatorname{Tr}\!\left(e^{-\tau D}\right)
\right]
\notag\\
&=
\frac{\ln(\Lambda/\mu_0)}{16\pi^{2}}
\int \text{d}^4x\,\sqrt{g}\,
\left[
\frac{11}{6}\gamma_{ab}
\mathbb{F}_{\mu\nu}^{a}\mathbb{F}^{b\mu\nu}
-\frac{13n}{180}R_{\mu\nu\rho\sigma}R^{\mu\nu\rho\sigma}
+\frac{22n}{45}R_{\mu\nu}R^{\mu\nu}
-\frac{5n}{36}R^{2}
\right]\notag\\
&\quad+\text{power-law divergences},
\end{align}
where $\mu_{0}$ is a reference scale. The coefficient of $\gamma_{ab}\mathbb{F}_{\mu\nu}^{a}\mathbb{F}^{b\mu\nu}$ contains the combined gauge field and Faddeev-Popov ghost contributions, while the curvature terms describe the response of the quadratic Yang-Mills sector to the background geometry.

Following the general prescription developed above, the contact part of the stress-tensor two-point function is obtained by taking two metric variations of $\mathcal{W}_{\Lambda}$ and is then contracted with the non-local kernel introduced in~\cite{Li:2025lpa}.  The resulting first-order deformed effective action is
\begin{align}
\mathcal{W}^{(1)}_{\Lambda}
=&
\frac{\lambda\ln(\Lambda/\mu_0)}{128\pi^4}\int \text{d}^4x\,\sqrt{g}\,
\Big[
\frac{11\Lambda^{2}}{2}
\gamma_{ab}\mathbb{F}_{\mu\nu}^{a}\mathbb{F}^{b\mu\nu}
\notag\\
&\quad+
n\Lambda^{2}\Big(
-\frac{39}{20}R_{\mu\nu\rho\sigma}R^{\mu\nu\rho\sigma}
+\frac{13}{6}R_{\mu\nu}R^{\mu\nu}
-\frac{1}{15}R^{2}
\Big)
\notag\\
&\quad+
\frac{11\ln(\Lambda/\mu_0)}{3}
\Big(
3R\,\gamma_{ab}\mathbb{F}_{\mu\nu}^{a}\mathbb{F}^{b\mu\nu}
-14R_{\mu\nu}\gamma_{ab}
\mathbb{F}^{a\mu}{}_{\alpha}\mathbb{F}^{b\nu\alpha}
\notag\\
&\quad
+3R_{\mu\nu\rho\sigma}\gamma_{ab}
\mathbb{F}^{a\mu\nu}\mathbb{F}^{b\rho\sigma}
\Big)
+O(R^3,\nabla^{6})
\Big].
\end{align}
As in the preceding examples, the terms proportional to
$\Lambda^{2}\ln(\Lambda/\mu_0)$ are local power-law divergences and can be absorbed into the corresponding counterterms.  After their subtraction, and upon discarding total derivatives on a compact manifold without boundary or under suitable boundary conditions, the logarithmically enhanced contribution becomes
\begin{equation}
\begin{aligned}
\mathcal{W}_{\mathrm{ren}}^{(1)}
=&
\frac{11\lambda\ln^2(\mu/\mu_0)}{384\pi^{4}}
\int \text{d}^4x\,\sqrt{g}\,
\Big[
3R\,\gamma_{ab}\mathbb{F}_{\mu\nu}^{a}\mathbb{F}^{b\mu\nu}
-14R_{\mu\nu}\gamma_{ab}
\mathbb{F}^{a\mu}{}_{\alpha}\mathbb{F}^{b\nu\alpha}
\\
&\quad+
3R_{\mu\nu\rho\sigma}\gamma_{ab}
\mathbb{F}^{a\mu\nu}\mathbb{F}^{b\rho\sigma}
\Big]
+O(R^3,\nabla^{6}).
\end{aligned}
\end{equation}

The induced action is therefore governed, at this derivative order, by non-minimal couplings between the background Yang-Mills field strength and spacetime curvature.  The relative coefficients receive contributions from both the gauge and ghost sectors and are compatible with the background-gauge-covariant formulation.  In contrast to the massive-vector example, the scheme-independent logarithmic sector contains no Einstein-Hilbert term.  Within the present massless perturbative setting, there is no intrinsic dimensionful parameter capable of supplying the coefficient of a two-derivative pure-gravity operator; the leading contribution instead consists of curvature-field-strength couplings.  This example thus shows explicitly that the geometric terms induced by the deformation depend not only on the field content of the seed theory, but also on its mass scales and gauge structure.

\section{Non-local stress tensor deformations of conformal field theories}\label{sec:cft}
In this section, the contribution of the non-contact, separated-point
part of the stress tensor two-point function - contained in the
ellipsis in (\ref{general seed theory stress tensor two-point function})
- to the effective gravitational action is examined.  For a CFT in
flat space, conformal invariance fixes the separated stress tensor
two-point function up to the normalization \(C_T\).  On a general curved
background, however, conformal symmetry does not determine the complete
short-distance singular structure of
\(\langle T_{\mu\nu}(x)T_{\rho\sigma}(y)\rangle_g\).  The universal
part is the leading identity contribution in the local \(TT\) OPE, while subleading singular terms are constrained by diffeomorphism and Weyl
Ward identities but generally depend on further CFT data, such as
\(TTT\) and \(TT\mathcal O\) OPE coefficients, anomaly coefficients,
background or state one-point functions, and local counterterm
conventions \cite{Osborn:1993cr,Erdmenger:1996yc,Hollands:2006hi,Bzowski:2017poo}.
In the first-order deformation of the effective action, UV divergences
from the near-coincidence region are controlled by these local
short-distance terms.  The coefficients of logarithmic divergences, or equivalently, genuinely non-local terms in the renormalized effective
action, are scheme independent, whereas power-law divergences and finite
local terms may be shifted by local gravitational counterterms.
\subsection{Stress tensor two-point function of CFT$_d$}
Let us start with a conformal field theory (CFT) in flat spacetime. When the background metric is the $d$-dimensional Euclidean metric, the stress tensor two-point function is uniquely fixed as \cite{Erdmenger:1996yc}
\begin{align}\label{flat spacetime two-point function}
    \langle{T_{\mu\nu}(x)T_{\rho\sigma}(y)}\rangle^{\text{CFT}}_{g=\eta}=\frac{C_T}{s^{2d}}I_{\mu\alpha}(s)I_{\nu\beta}(s)\mathcal{E}^{\alpha\beta}_{\rho\sigma},
\end{align}
where $s=x-y$ and $I_{\mu\nu}(s)=\eta_{\mu\nu}-2s_{\mu}s_{\nu}/s^2$. $\mathcal{E}^{\alpha\beta}_{\rho\sigma}$ is the projector onto symmetric traceless tensors, $\mathcal{E}^{\alpha\beta}_{\rho\sigma}=(1/2)(\delta^{\alpha}_{\rho}\delta^{\beta}_{\sigma}+\delta^{\alpha}_{\sigma}\delta^{\beta}_{\rho})-(1/d)\eta^{\alpha\beta}\eta_{\rho\sigma}$.\par
For a \(d\)-dimensional CFT on a general curved background, the full
separated-point stress tensor two-point function is not fixed by
conformal symmetry alone.  The appropriate short-distance description is
the local, covariant \(TT\) OPE. Its leading identity contribution is
universal and is obtained by covariantizing the flat-space tensor
structure fixed by \(C_T\),
\begin{align}\label{two-point function leading order}
    \langle{T_{\mu\nu}(x)T_{\rho\sigma}(y)}\rangle^{\text{CFT}}_{g}=\frac{\mathcal{T}_{\mu\nu,\rho\sigma}(x,y)}{(2\sigma(x,y))^{d}}+O(\sigma^{-d+1}).
\end{align}
$\mathcal{T}_{\mu\nu,\rho\sigma}(x,y)$ can be obtained through a covariant generalization of the flat-spacetime two-point function (\ref{flat spacetime two-point function}),
\begin{align}\label{total two-point coefficient}
    \mathcal{T}_{\mu\nu,\rho\sigma}(x,y)=C_T\Big[\frac{1}{2}\Big(\mathcal{I}_{\mu\rho}(x,y)\mathcal{I}_{\nu\sigma}(x,y)+\mathcal{I}_{\mu\sigma}(x,y)\mathcal{I}_{\nu\rho}(x,y)\Big)-\frac{1}{d}g_{\mu\nu}(x)g_{\rho\sigma}(y)\Big],
\end{align}
where $\mathcal{I}_{\mu\rho}(x,y)=g_{\mu\rho}(x,y)-\sigma_{\mu}\sigma_{\rho}/\sigma$. The subleading terms, however, are not
universal: they depend on the operator spectrum and OPE data of the CFT,
on possible one-point functions in the chosen background or state, and
on the choice of local counterterm scheme.\par
Hadamard parametrices are nevertheless useful in a more restricted
setting.  For a specified free field theory, or for a perturbative
Lagrangian theory, after choosing its field content, the singular part of the fundamental propagators may be constructed by the heat-kernel expansion.  Applying the appropriate point-split stress-tensor differential operator to those propagators then gives the singular part of \(\langle T_{\mu\nu}(x)T_{\rho\sigma}(y)\rangle\) for that particular theory \cite{Radzikowski:1996pa,Sahlmann:2000zr,Hollands:2001fb,Moretti:2001qh,Decanini:2005eg}. This singular structure can only depend on the local geometry, and thus can be expressed naturally in terms of the Synge world function $\sigma(x,y)$, the parallel propagator $g_{\mu}{}^{\rho}(x,y)$, and local geometric invariants. For even spacetime dimension, the general form of the stress tensor two-point function is given by the Hadamard parametrix \cite{Radzikowski:1996pa,Fewster:2013lqa,Moretti:2021pzz} as follows,
\begin{align}\label{general form of two-point function}
    \langle{T_{\mu\nu}(x)T_{\rho\sigma}(y)}\rangle^{\text{Had}}_{g}=\frac{U_{\mu\nu,\rho\sigma}(x,y)}{(2\sigma(x,y))^{d}}+V_{\mu\nu,\rho\sigma}(x,y)\ln(\sigma(x,y))+W_{\mu\nu,\rho\sigma}(x,y).
\end{align}
Here $U_{\mu\nu,\rho\sigma}(x,y)$, $V_{\mu\nu,\rho\sigma}(x,y)$, and $W_{\mu\nu,\rho\sigma}(x,y)$ are smooth functions. $U_{\mu\nu,\rho\sigma}(x,y)$ and $V_{\mu\nu,\rho\sigma}(x,y)$ are constructed from geometric quantities and are uniquely determined by the Hadamard recursion relations \cite{Decanini:2005eg,Kay:1988mu,DeWitt:1960fc,garabedian1964partial}. $W_{\mu\nu,\rho\sigma}(x,y)$ encodes detailed information about the quantum state, including the global topology of the system and long-range correlations.\par
As a self-consistency check, we verify that, for $x \neq y$, the leading-order two-point function (\ref{two-point function leading order}) satisfies the tracelessness condition of the stress tensor in CFT. Contracting the product $\mathcal{I}_{\mu\rho}(x,y)\mathcal{I}_{\nu\sigma}(x,y)$ with $g^{\mu\nu}(x)$ and using the properties of the Synge world function and the parallel propagator yields
\begin{align}
    g^{\mu\nu}(x)\mathcal{I}_{\mu\rho}(x,y)\mathcal{I}_{\nu\sigma}(x,y)&=g^{\mu\nu}(x)\Big(g_{\mu\rho}(x,y)-\frac{\sigma_{\mu}\sigma_{\rho}}{\sigma}\Big)\Big(g_{\nu\sigma}(x,y)-\frac{\sigma_{\nu}\sigma_{\sigma}}{\sigma}\Big)\notag\\
    &=g_{\rho\sigma}(y)-\frac{\sigma_{\sigma}\sigma_{\rho}}{\sigma}-\frac{\sigma_{\rho}\sigma_{\sigma}}{\sigma}+\frac{2\sigma\sigma_{\rho}\sigma_{\sigma}}{\sigma^2}\notag\\
    &=g_{\rho\sigma}(y).
\end{align}
Plugging it and (\ref{total two-point coefficient}) into the trace of (\ref{two-point function leading order}), one obtains
\begin{align}
    g^{\mu\nu}(x)\frac{\mathcal{T}_{\mu\nu,\rho\sigma}(x,y)}{(2\sigma(x,y))^{d}}=0,\quad\quad\text{for }x\neq y.
\end{align}
Similarly, conservation of the stress tensor two-point function is verified by computing its covariant derivative at leading order. By employing the Riemann normal coordinates, which will be introduced in the next subsection, we can show that the divergence of (\ref{two-point function leading order}) is subleading,
\begin{align}
    \nabla^{\mu}_{x}\frac{\mathcal{T}_{\mu\nu,\rho\sigma}(x,y)}{(2\sigma(x,y))^{d}}=O(\sigma^{-d+1}).
\end{align}\par
For later convenience, the leading-order coefficient (\ref{total two-point coefficient}) is decomposed into two parts as follows,
\begin{equation}
\begin{aligned}\label{decomposition of Tmunurhosigma}
    \mathcal{T}_{\mu\nu,\rho\sigma}(x,y)&\equiv\mathcal{T}^{(1)}_{\mu\nu,\rho\sigma}(x,y)+\mathcal{T}^{(2)}_{\mu\nu,\rho\sigma}(x,y),\\
    \mathcal{T}^{(1)}_{\mu\nu,\rho\sigma}(x,y)&=C_T\Big(g_{\mu(\rho}(x,y)g_{\sigma)\nu}(x,y)-\frac{1}{d}g_{\mu\nu}(x)g_{\rho\sigma}(y)\Big),\\
    \mathcal{T}^{(2)}_{\mu\nu,\rho\sigma}(x,y)&=-\frac{C_T}{\sigma}\Big(\sigma_{\mu}g_{\nu(\rho}(x,y)\sigma_{\sigma)}+\sigma_{\nu}g_{\mu(\rho}(x,y)\sigma_{\sigma)}-\frac{\sigma_{\mu}\sigma_{\nu}\sigma_{\rho}\sigma_{\sigma}}{\sigma}\Big).
\end{aligned}
\end{equation}
The first part is well defined in the coincidence limit $y\to x$, whereas for the second part, however, a unique limit is not obtained, and its value is determined by the direction of approach.
\subsection{Dimensional regularization}
Next, we consider the leading-order contribution of the two-point function to the first-order correction of the effective action, and compute its UV logarithmic divergences. For the subleading and higher orders of the two-point function, once their explicit forms are determined through the Hadamard recursion relations, a similar analysis can be carried out. By plugging (\ref{two-point function leading order}) into (\ref{flow equation for effective action}), we have
\begin{align}\label{1st effective action from non-contact term}
    \mathcal{W}^{(1)}[g]=\int \text{d}^dx\text{d}^dy\sqrt{g(x)g(y)}\,\frac{H^{\mu\nu,\rho\sigma}(x,y)\mathcal{T}_{\mu\nu,\rho\sigma}(x,y)}{(2\sigma(x,y))^{d}}.
\end{align}
In the following, we select a specific class of non-local kernels to evaluate this integral; the treatment for more general non-local kernels is analogous. Motivated by the linearized graviton propagator, we set $G^{\mu\nu,\alpha\beta}(x,y)$ as the Green's function of the Laplace-type operator,
\begin{align}
    \Big(-g_{\mu\rho}(x)g_{\nu\sigma}(x)\Box_x+N_{\mu\nu\rho\sigma}(x)\Big)G^{\mu\nu,\alpha\beta}(x,y)=\tilde\delta(x-y)\delta_{\rho}^{\alpha}\delta_{\sigma}^{\beta}.
\end{align}
By employing the Mellin transformation, this Green's function can be expressed in terms of the heat kernel of the Laplace-type operator as
\begin{align}
    G^{\mu\nu,\alpha\beta}(x,y)=\int_{0}^{\infty} \text{d}\tau\,K^{\mu\nu,\alpha\beta}(\tau;x,y;\Delta),
\end{align}
while the heat kernel has the following small-$\tau$ expansion,
\begin{align}
    K^{\mu\nu,\alpha\beta}(\tau;x,y;\Delta)=\frac{\Delta^{1/2}(x,y)}{(4\pi\tau)^{d/2}}e^{-\sigma(x,y)/2\tau}\sum_{n=0}^{\infty}a^{\mu\nu,\alpha\beta}_{n}(x,y)\tau^{n}.
\end{align}
We decompose the total Green's function into the following two parts,
\begin{align}
    G^{\mu\nu,\alpha\beta}(x,y)=G^{\mu\nu,\alpha\beta}_{\text{local}}(x,y)+G^{\mu\nu,\alpha\beta}_{\text{global}}(x,y).
\end{align}
The first part arises from the integration of the heat kernel expansion at the UV region, which can be formally written as
\begin{align}\label{local part of the Green's function}
    G^{\mu\nu,\alpha\beta}_{\text{local}}(x,y)=\frac{\Delta^{1/2}(x,y)}{(4\pi)^{d/2}}\sum_{n=0}^{\infty}a^{\mu\nu,\alpha\beta}_{n}(x,y)I_{n}(x,y).
\end{align}
Here, the integral is given by $I_{n}=\int_{0}^{L^2}\text{d}\tau\,e^{-\sigma/2\tau}\tau^{n-d/2}$, in which the upper limit $L^2$ is an arbitrary scale introduced to exclude the contribution from IR divergence. For simplicity, we employ the dimensional regularization throughout, $d\mapsto d-\varepsilon$. The regularized $I_n$ can be expressed as 
\begin{align}\label{integral In}
    I_{n}^{(\varepsilon)}&=\mu^{\varepsilon}\Big(\frac{\sigma}{2}\Big)^{k_n+\varepsilon/2}\Gamma\Big(-k_n-\frac{\varepsilon}{2},\frac{\sigma}{2L^2}\Big)\notag\\
    &=\mu^{\varepsilon}\Big(\frac{\sigma}{2}\Big)^{k_n+\varepsilon/2}\Gamma\Big(-k_n-\frac{\varepsilon}{2}\Big)-\mu^{\varepsilon}\sum_{m=0}^{\infty}\frac{(-1)^mL^{2k_n-2m+\varepsilon}}{m!(m-k_n-\varepsilon/2)}\Big(\frac{\sigma}{2}\Big)^m.
\end{align}
where $k_n=n+1-d/2$. The second part, $G^{\mu\nu,\alpha\beta}_{\text{global}}(x,y)$, represents the smooth remainder term, which depends on global spectral data and is sensitive to the specific boundary conditions and global topology. This part cannot be expressed in terms of the heat kernel coefficients $\{a^{\mu\nu,\rho\sigma}_n(x,y)\}$, and therefore its contribution to the effective gravitational action cannot be evaluated using the methods developed in this paper\footnote{Importantly, the smooth remainder does not contribute to the UV divergences of the Green's function. However, it does contribute to the UV divergences of the integral (\ref{1st effective action from non-contact term}), due to the negative-power prefactor $(2\sigma)^{-d}$ in the stress-tensor two-point function. Consequently, the smooth remainder cannot be neglected when considering the UV logarithmic divergences that contribute to the effective gravitational action after renormalization.}. Plugging (\ref{local part of the Green's function}) and (\ref{integral In}) into (\ref{1st effective action from non-contact term}), and decomposing the integral in powers of $\sigma(x,y)$ as follows
\begin{align}
    \mathcal{W}_{\text{local}}^{(1)}[g]&=\sum_{n=0}^{\infty}\Big(\mathcal{U}_{n}[g]+\sum_{m=0}^{\infty}\mathcal{V}_{m,n}[g]\Big),
\end{align}
in which $\{\mathcal{U}_n\}$ are defined as
\begin{align}\label{Un before renormalization}
    \mathcal{U}_{n}&=\int \text{d}\mu(x)\int \text{d}\mu(y)\Delta(x,y)\mathcal{J}^{[\varepsilon]}_n(x,y)\mu^{2\varepsilon}\Gamma\Big(-k_n-\frac{\varepsilon}{2}\Big)(2\sigma)^{k_n+\varepsilon/2-D},
\end{align}
and $\{\mathcal{V}_{m,n}\}$ are defined as
\begin{align}\label{Vn before renormalization}
    \mathcal{V}_{m,n}&=-\int \text{d}\mu(x)\int \text{d}\mu(y)\Delta(x,y)\mathcal{J}^{[\varepsilon]}_n(x,y)\mu^{2\varepsilon}\frac{(-1)^m(2L)^{2k_n-2m+\varepsilon}}{m!(m-k_n-\varepsilon/2)}(2\sigma)^{m-D}.
\end{align}
Here $D=d-\varepsilon$. And the biscalar $\mathcal{J}^{[\varepsilon]}_n(x,y)$ takes the form
\begin{equation}
    \mathcal{J}^{[\varepsilon]}_n(x,y)=\frac{\Delta^{-1/2}(x,y)}{2^{2k_n+\varepsilon}(4\pi)^{D/2}}\mathcal{T}_{\mu\nu,\rho\sigma}(x,y)\tilde{a}^{\mu\nu,\rho\sigma}_{n}(x,y),
\end{equation}
where
\begin{equation}
    \tilde{a}^{\mu\nu,\rho\sigma}_{n}(x,y)\equiv I^{\mu\nu}_{\alpha\beta}(x)a^{\alpha\beta,\gamma\delta}_{n}(x,y)J^{\rho\sigma}_{\gamma\delta}(y).
\end{equation}
We can decompose $\mathcal{J}^{[\varepsilon]}_n(x,y)$ into the following parts, depending on whether they are well defined in the coincidence limit,
\begin{align}\label{decomposition of Jn}
\mathcal{J}^{[\varepsilon]}_n(x,y)=\bar{\mathcal{J}}^{[\varepsilon]}_n(x,y)+\frac{\sigma_{\mu}\sigma_{\rho}}{2\sigma}\bar{\mathcal{J}}^{[\varepsilon]\mu\rho}_n(x,y)+\frac{\sigma_{\mu}\sigma_{\nu}\sigma_{\rho}\sigma_{\sigma}}{4\sigma^2}\bar{\mathcal{J}}^{[\varepsilon]\mu\nu,\rho\sigma}_n(x,y).
\end{align}
Here $\bar{\mathcal{J}}^{[\varepsilon]}_n(x,y)$ denotes the part contributed by $\mathcal{T}^{(1)}_{\mu\nu,\rho\sigma}(x,y)$ in (\ref{decomposition of Tmunurhosigma}), $\bar{\mathcal{J}}^{[\varepsilon]\mu\rho}_n(x,y)$ and $\bar{\mathcal{J}}^{[\varepsilon]\mu\nu,\rho\sigma}_n(x,y)$ are given by
\begin{align}
    \bar{\mathcal{J}}^{[\varepsilon]\mu\rho}_n(x,y)&=-\frac{4C_{T}\Delta^{-1/2}(x,y)}{2^{2k_n+\varepsilon}(4\pi)^{D/2}}g_{\nu\sigma}(x,y)\tilde{a}_{n}^{(\mu\nu),(\rho\sigma)}(x,y),\\
    \bar{\mathcal{J}}^{[\varepsilon]\mu\nu,\rho\sigma}_n(x,y)&=\frac{4C_{T}\Delta^{-1/2}(x,y)}{2^{2k_n+\varepsilon}(4\pi)^{D/2}}\tilde{a}_{n}^{\mu\nu,\rho\sigma}(x,y).
\end{align}
To perform the integration over $y$, it is convenient to introduce the Riemann normal coordinates \cite{Poisson:2011nh,Brewin:2009se} as follows
\begin{align}
    \xi^{a}=-e^{a}{}_{\mu}(x)\sigma^{\mu}(x,y),
\end{align}
where $e^{a}{}_{\mu}(x)$ is the vielbein. In this coordinate system, the Synge world function can be simply expressed as $2\sigma(x,y)=\eta_{ab}\xi^{a}\xi^{b}$. The metric can be expanded in powers of $\xi^{a}$ as
\begin{align}
    g_{ab}(\xi)=\eta_{ab}-\frac{1}{3}R_{acbd}(x)\xi^{c}\xi^{d}+O(\xi^c\xi^{d}\xi^{e}).
\end{align}
The determinant of the metric is simply given by
\begin{align}
    \sqrt{g(\xi)}=\Delta^{-1}(x,y(\xi))
\end{align}
The expansion of $\bar{\mathcal{J}}_n(x,y)$ can be formally written as \cite{Poisson:2011nh}
\begin{align}\label{decomposition of Jn 1}
    \bar{\mathcal{J}}^{[\varepsilon]}_n(x,y(\xi))&=[\bar{\mathcal{J}}^{[\varepsilon]}_n]+\Big([\nabla^{x}_{\mu}\bar{\mathcal{J}}^{[\varepsilon]}_n]-\nabla^{x}_{\mu}[\bar{\mathcal{J}}^{[\varepsilon]}_n]\Big)\sigma^{\mu}+O(\sigma^{\mu}\sigma^{\nu})\notag\\
    &=[\bar{\mathcal{J}}^{[\varepsilon]}_n]-\Big([\nabla^{x}_{\mu}\bar{\mathcal{J}}^{[\varepsilon]}_n]-\nabla^{x}_{\mu}[\bar{\mathcal{J}}^{[\varepsilon]}_n]\Big)e_{a}{}^{\mu}\xi^{a}+O(\xi^a\xi^b)\notag\\
    &=\sum_{l=0}^{\infty}\bar{\mathcal{J}}^{[\varepsilon]}_{na_1\cdots a_l}(x)\xi^{a_1}\cdots\xi^{a_l},
\end{align}
where $[\bar{\mathcal{J}}^{[\varepsilon]}_n](x)=\lim_{y\to x}\bar{\mathcal{J}}^{[\varepsilon]}_n(x,y)$ denotes its coincidence limit. The other two coefficients in (\ref{decomposition of Jn}), $\bar{\mathcal{J}}^{[\varepsilon]\mu\rho}_n(x,y)$ and $\bar{\mathcal{J}}^{[\varepsilon]\mu\nu,\rho\sigma}_n(x,y)$, admit similar expansions,
\begin{align}\label{decomposition of Jn 2}
    \bar{\mathcal{J}}^{[\varepsilon]\mu\rho}_n(x,y(\xi))&=e_{a}{}^{\mu}(x)e_{c}{}^{\rho}(y)\sum_{l=0}^{\infty}\bar{\mathcal{J}}^{[\varepsilon]ac}_{na_1\cdots a_l}(x)\xi^{a_1}\cdots\xi^{a_l},\notag\\
    \bar{\mathcal{J}}^{[\varepsilon]\mu\nu,\rho\sigma}_n(x,y(\xi))&=e_{a}{}^{\mu}(x)e_{b}{}^{\nu}(x)e_{c}{}^{\rho}(y)e_{d}{}^{\sigma}(y)\sum_{l=0}^{\infty}\bar{\mathcal{J}}^{[\varepsilon]abcd}_{na_1\cdots a_l}(x)\xi^{a_1}\cdots\xi^{a_l}.
\end{align}
Next, we switch to the hyperspherical coordinates,
\begin{align}
    \xi^{a}=r \hat n^{a},
\end{align}
where $\hat{n}^a(\Omega)$ is the unit vector on the $(D-1)$-sphere. The coordinate volume element transforms as $\text{d}^D\xi=r^{D-1}\text{d}r\text{d}\Omega_{D-1}$. By plugging this coordinate transformation into (\ref{decomposition of Jn}), (\ref{decomposition of Jn 1}), and (\ref{decomposition of Jn 2}), we obtain
\begin{align}
    \mathcal{J}^{[\varepsilon]}_n(x,r,\Omega)=\sum_{l=0}^{\infty}\Big[\bar{\mathcal{J}}^{[\varepsilon]}_{n}+\hat n_{a}\hat n_{c}\bar{\mathcal{J}}^{[\varepsilon]ac}_{n}+\hat n_{a}\hat n_{b}\hat n_{c}\hat n_{d}\bar{\mathcal{J}}^{[\varepsilon]abcd}_{n}\Big]_{a_1\cdots a_l}(x)r^l\hat n^{a_1}\cdots\hat n^{a_l}.
\end{align}

\par
One can further plug the above equations into $\{\mathcal{U}_n\}$ and $\{\mathcal{V}_{m,n}\}$ and perform the integration over $\xi$. However, this procedure is plagued by two obvious ambiguities. The first ambiguity is that the radial coordinate $r$ is integrated over $(0, \infty)$, whereas the Riemann normal coordinates are only well-defined in a neighborhood of the base point $x$. For a general curved background, such a neighborhood does not necessarily extend to infinity. The second ambiguity concerns the radial integration, which involves a scaleless integral of the form $\int_0^\infty \text{d}r \, r^{\alpha}$. In dimensional regularization, such a scaleless integral vanishes due to a cancellation between UV and IR divergences \cite{Abreu:2022mfk, El-Menoufi:2015cqw, Schwartz:2014sze}. Nevertheless, in our previous analysis, we employed a small-$\tau$ expansion of the Green's function and considered only the contribution from $G^{\mu\nu,\rho\sigma}_{\text{local}}$. By construction, this part captures the UV behavior of the Green's function solely, whereas its IR behavior is encoded in $G^{\mu\nu,\rho\sigma}_{\text{global}}$, which has not been taken into account. To avoid these ambiguities, we introduce an arbitrary scale $R$ to divide the UV and IR regions \cite{Schwartz:2014sze}. Since only the UV divergent part of the effective action is of interest, the $r$-integration region is restricted to $(0, R)$. Putting everything together, we can express the coefficients $\{\mathcal{U}_n\}$ in (\ref{Un before renormalization}) as follows,
\begin{align}
    \mathcal{U}_n&=\int \text{d}\mu(x)\sum_{l=0}^{\infty}\Big[\bar{\mathcal{J}}^{[\varepsilon]}_{na_1\cdots a_l}\Omega^{a_1\cdots a_l}+\bar{\mathcal{J}}^{[\varepsilon]ac}_{na_1\cdots a_l}\Omega^{a_1\cdots a_l}_{ac}+\bar{\mathcal{J}}^{[\varepsilon]abcd}_{na_1\cdots a_l}\Omega^{a_1\cdots a_l}_{abcd}\Big]J^{[\varepsilon,R]}_{l,n}.
\end{align}
Here $J^{[\varepsilon, R]}_{l,n}$ represents the radial integration, which can be evaluated via analytic continuation,
\begin{align}
    &J^{[\varepsilon,R]}_{l,n}=\int_{0}^{R} \text{d}r\,\mu^{2\varepsilon}\Gamma\Big(-k_n-\frac{\varepsilon}{2}\Big)r^{l+2n+1-2D}\notag\\
    &=
    \begin{cases}
    (-k_n-1)!\frac{R^{q_{l,n}}}{q_{l,n}}+O(\varepsilon)&(n
    \leq\frac{d}{2}-2, l\neq d-2k_n),\\
   (-k_n-1)!\Big(\frac{1}{2\varepsilon}+\ln\mu R-\frac{1}{4}\psi(-k_n)\Big)+O(\varepsilon)&(n
    \leq\frac{d}{2}-2, l=d-2k_n),\\
    \frac{(-1)^{k_n+1}}{k_n!}R^{q_{l,n}}\Big(\frac{2}{q_{l,n}\varepsilon}+\frac{4\ln\mu R-\psi(k_n+1)}{q_{l,n}}-\frac{4}{q_{l,n}^2}\Big)+O(\varepsilon)&(n
    \geq\frac{d}{2}-1, l\neq d-2k_n),\\
    \frac{(-1)^{k_n+1}}{k_n!}\Big(\frac{1}{\varepsilon^2}+\frac{4\ln\mu R-\psi(k_n+1)}{2\varepsilon}-\psi(k_n+1)\ln\mu R\\
    \quad\quad\quad\quad+2\ln^2\mu R+\frac{\psi^2(k_n+1)+\psi'(k_n+1)}{8}\Big)+O(\varepsilon)&(n
    \geq\frac{d}{2}-1, l=d-2k_n),
\end{cases}
\end{align}
where $q_{l,n}=l+2n+2-2d$ and $\psi(k_n+1)=-\gamma_{E}+\sum_{j=1}^{k_n}1/j$ is the digamma function. And the contribution from the angular integration is denoted as
\begin{equation}
\Omega^{a_1\dots a_l} \;\equiv\; \int_{\mathbb{S}^{D-1}} \text{d}\Omega \;\hat{n}^{a_1}\hat{n}^{a_2}\cdots \hat{n}^{a_l},
\end{equation}
where $\text{d}\Omega$ denotes the standard volume element on the unit sphere $\mathbb{S}^{D-1}$. The result is given by \cite{le2018anomalous} \footnote{As an example, for $l=4$, Eq.~\eqref{angular-integral-general} gives the rank-four angular moment
\begin{equation}
\Omega^{a_1a_2a_3a_4}=
\frac{\Omega_{D-1}}{D(D+2)}
\left(
\eta^{a_1a_2}\eta^{a_3a_4}
+\eta^{a_1a_3}\eta^{a_2a_4}
+\eta^{a_1a_4}\eta^{a_2a_3}
\right).
\end{equation}}
\begin{equation}\label{angular-integral-general}
    \Omega^{a_1\cdots a_l}=
     \begin{cases}
  0,&\text{for odd } l,\\
   \frac{\Omega_{D-1}}{D(D+2)\cdots (D+l-2)}\Big(\eta^{a_1a_2}\cdots\eta^{a_{l-1}a_{l}}+\text{pairings}\Big),&\text{for even } l.
\end{cases}
\end{equation}
Here the volume of $\mathbb{S}^{D-1}$ is $\Omega_{D-1}=\frac{2\pi^{D/2}}{\Gamma(D/2)}$. Substituting these results back into $\mathcal{U}_n$ and expanding it in powers of $\varepsilon$ yields the regularized effective action. Similarly, the coefficients $\{\mathcal{V}_{m,n}\}$ in (\ref{Vn before renormalization}) can be evaluated as follows,
\begin{align}
    \mathcal{V}_{m,n}=-\int \text{d}\mu(x)\sum_{l=0}^{\infty}\Big[\bar{\mathcal{J}}^{[\varepsilon]}_{na_1\cdots a_l}\Omega^{a_1\cdots a_l}+\bar{\mathcal{J}}^{[\varepsilon]ac}_{na_1\cdots a_l}\Omega^{a_1\cdots a_l}_{ac}+\bar{\mathcal{J}}^{[\varepsilon]abcd}_{na_1\cdots a_l}\Omega^{a_1\cdots a_l}_{abcd}\Big]K^{[\varepsilon,L,R]}_{l,m,n}.
\end{align}
The radial integral $K^{[\varepsilon,R]}_{l,m,n}$ can be evaluated via analytic continuation,
\begin{align}
    K^{[\varepsilon,R]}_{l,m,n}&=\int_{0}^{R}\text{d}r\,\mu^{2\varepsilon}\frac{(-1)^m(2L)^{2k_n-2m+\varepsilon}}{m!(m-k_n-\varepsilon/2)}r^{l+2m-1-D}\notag\\
    &=
    \begin{cases}
  \frac{(-1)^m(2L)^{2k_n-2m}R^{l+2m-d}}{m!(m-k_n)(l+2m-d)}+O(\varepsilon)&(l\neq d-2m,m
    \neq k_n),\\
   \frac{(-1)^m(2L)^{2k_n-2m}}{m!(m-k_n)}\Big(\frac{1}{\varepsilon}+\ln(2\mu^2LR)+\frac{1}{2(m-k_n)}\Big)+O(\varepsilon)&(l=d-2m,m
    \neq k_n),\\
    \frac{(-1)^{k_n+1}2R^{l+2k_n-d}}{k_n!(l+2k_n-d)}\Big(\frac{1}{\varepsilon}+\ln(2\mu^2LR)-\frac{1}{l+2k_n-d}\Big)+O(\varepsilon)&(l\neq d-2m,m
    =k_n),\\
    \frac{(-1)^{k_n+1}}{k_n!}\Big(\frac{2}{\varepsilon^2}+\frac{2\ln(2\mu^2LR)}{\varepsilon}+\ln^2(2\mu^2LR)\Big)+O(\varepsilon)&(l= d-2m,m=k_n).
\end{cases}
\end{align}
\subsection{Scheme-independent effective gravitational action}
The divergent terms of the form $1/\varepsilon$ and $1/\varepsilon^2$ in $\{\mathcal{U}_{n}\}$ and $\{\mathcal{V}_{m,n}\}$ are canceled by introducing local counterterms, and the remaining part gives the renormalized effective action. The renormalized result depends explicitly on the arbitrary scales $(L, R)$, and is therefore scheme-dependent. This is, in fact, expected, as we have artificially separated the IR contribution. When the IR renormalization effects are also taken into account, the full renormalized effective action should be independent of the unphysical parameters $(L, R)$,
 \begin{align}
     \mathcal{U}_n^{\text{ren}}&=\mathcal{U}_n^{\text{ren,UV}}(R)+\mathcal{U}_n^{\text{ren,IR}}(R),\quad \frac{d}{dR}\mathcal{U}_n^{\text{ren}}=0,\notag\\
     \mathcal{V}_{m,n}^{\text{ren}}&=\mathcal{V}_{m,n}^{\text{ren,UV}}(L,R)+\mathcal{V}_{m,n}^{\text{ren,IR}}(L,R),\quad \frac{\partial}{\partial L}\mathcal{V}_{m,n}^{\text{ren}}=\frac{\partial}{\partial R}\mathcal{V}_{m,n}^{\text{ren}}=0.
 \end{align}
In addition, the constant finite terms in the renormalized effective action are also scheme-dependent. Only the terms of order $\ln\mu$ and $\ln^2\mu$ in the regularized effective action ultimately give scheme-independent (SI) finite contributions. By introducing a reference scale $\mu_0$ to extract these terms, one obtains the scheme-independent coefficients $\{\mathcal{U}^{\text{SI}}_{n}\}$,
\begin{equation}
\begin{aligned}
    \mathcal{U}_{\frac{d}{2}-1>n\geq0}^{\text{SI}}&=\int \text{d}\mu(x)(-k_n-1)!\ln(\mu/\mu_0)\mathcal{J}^{[0]}_{n,-q_{0,n}}(x),\\
    \mathcal{U}_{d-1\geq n\geq \frac{d}{2}-1}^{\text{SI}}&=\int \text{d}\mu(x)\frac{(-1)^{k_n}}{k_n!}\Big(\psi(k_n+1)\ln(\mu/\mu_0)-2\ln^2(\mu/\mu_0)\Big)\mathcal{J}^{[0]}_{n,-q_{0,n}}(x),
\end{aligned}
\end{equation}
and the scheme-independent coefficients $\{\mathcal{V}^{\text{SI}}_{m,n}\}$,
\begin{align}
        \mathcal{V}^{\text{SI}}_{k_n,d-1\geq n\geq\frac{d}{2}-1}&=\int \text{d}\mu(x)\,\frac{4(-1)^{k_n}}{k_n!}\ln^2(\mu/\mu_0)\mathcal{J}^{[0]}_{n,-q_{0,n}}(x),
\end{align}
Here we have used $\mathcal{J}^{[\varepsilon]}_{n,l}\equiv\bar{\mathcal{J}}^{[\varepsilon]}_{na_1\cdots a_l}\Omega^{a_1\cdots a_l}+\bar{\mathcal{J}}^{[\varepsilon]ac}_{na_1\cdots a_l}\Omega^{a_1\cdots a_l}_{ac}+\bar{\mathcal{J}}^{[\varepsilon]abcd}_{na_1\cdots a_l}\Omega^{a_1\cdots a_l}_{abcd}$ to simplify the notation. As a simple example, we calculate the explicit form of the $\mathcal{U}^{\text{SI}}_{d-1}$ term in four dimensions. We take the simplest non-local kernel, 
$I^{\mu\nu}_{\alpha\beta}=J^{\mu\nu}_{\alpha\beta}=\delta^{\mu}_{\alpha}\delta^{\nu}_{\beta}$ and $N_{\mu\nu\rho\sigma}=0$, 
and employ the Seeley–DeWitt coefficients for the heat kernel of a Laplace-type operator \cite{Vassilevich:2003xt} to obtain
\begin{align}
    \mathcal{U}_{3}^{\text{SI}}&=\frac{(\frac{3}{2}-\gamma_{E})\ln\frac{\mu}{\mu_0}-2\ln^2\frac{\mu}{\mu_0}}{768}C_{T}\int \text{d}^4x\sqrt{g}\Big[g_{\mu\rho}g_{\nu\sigma}+g_{\mu\sigma}g_{\nu\rho}-\frac{1}{2}g_{\mu\nu}g_{\rho\sigma}\Big][\tilde{a}^{\mu\nu,\rho\sigma}_{3}]\notag\\
    &=\frac{(\frac{3}{2}-\gamma_{E})\ln\frac{\mu}{\mu_0}-2\ln^2\frac{\mu}{\mu_0}}{60480}C_T\int \text{d}^4x\sqrt{g}\Big(18\Box^2R+11R\Box R-6R_{\mu\nu}\Box R^{\mu\nu}\notag\\
    &\quad+3R_{\mu\nu\rho\sigma}\Box R^{\mu\nu\rho\sigma}+28R_{\mu\nu}\nabla^{\nu}\nabla_{\rho}R^{\mu\rho}+\frac{35}{9}R^3-\frac{14}{3}RR_{\mu\nu}R^{\mu\nu}+\frac{14}{3}RR_{\mu\nu\rho\sigma}R^{\mu\nu\rho\sigma}\notag\\
    &\quad-\frac{172}{9}R_{\mu\nu}R^{\mu\rho}R^{\nu}{}_{\rho}-\frac{76}{3}R_{\mu\nu}R_{\rho\sigma}R^{\mu\rho\nu\sigma}-\frac{16}{3}R_{\mu\nu}R^{\mu\rho\sigma\delta}R^{\nu}{}_{\rho\sigma\delta}-\frac{44}{9}R_{\mu\nu\rho\sigma}R^{\mu\nu\delta\gamma}R^{\rho\sigma}{}_{\delta\gamma}\notag\\
    &\quad-\frac{80}{9}R_{\mu\nu\rho\sigma}R^{\mu\delta\rho\gamma}R^{\nu}{}_{\delta}{}^{\sigma}{}_{\gamma}\Big).
\end{align}
Here we have omitted the boundary terms. To make the effective gravitational action depend explicitly on low-order curvature terms, a mass term can be added to the Green function equation for the non-local kernel, with the simplest choice being $N_{\mu\nu\rho\sigma}(x)=g_{\mu(\rho}(x)g_{\sigma)\nu}(x)m^{2}$. The resulting $\mathcal{U}^{\text{SI}}_{d-1}$ term in four dimensions takes the form
\begin{align}
    \mathcal{U}_{3}^{\text{SI}}(m)&=\frac{(\frac{3}{2}-\gamma_{E})\ln\frac{\mu}{\mu_0}-2\ln^2\frac{\mu}{\mu_0}}{4320}C_T\int \text{d}^4x\sqrt{g}\Big[-60m^6+30m^4R-12m^2\Box R\notag\\
    &\quad-5m^2R^2+2m^2R_{\mu\nu}R^{\mu\nu}-2m^2R_{\mu\nu\rho\sigma}R^{\mu\nu\rho\sigma}\Big]+\mathcal{U}_{3}^{\text{SI}}(0).
\end{align}
\section{Effective gravitational action from the conformal anomaly}\label{sec:anomaly}
In the previous section, we derived the first-order effective gravitational action generated by a non-local deformation built from two stress tensors. We now consider its trace-channel counterpart, obtained by replacing the two insertions by 
\begin{align}
    \Theta=g_{\mu\nu}T^{\mu\nu}.
\end{align} 
For a CFT, the non-contact part of the two-point function $\langle{\Theta\Theta}\rangle$ vanishes. The remaining contribution is purely contact and is fixed by the conformal Weyl anomaly. This observation makes the trace deformation a particularly economical setting for obtaining the first-order effective gravitational action from anomaly data. \par
The connection between trace dynamics and induced gravity was already emphasized in the Adler-Zee approach \cite{Adler:1980a, Adler:1980bx, Zee:1981mk, Zee:1982id, Brown:1983zz}. In that framework, a microscopic quantum field theory is coupled to an external metric and the matter fields are integrated out, leading to a metric functional \(\mathcal{W}_{\rm m}[g]\). Adler and Zee isolated this coefficient by probing the theory with a purely scalar metric perturbation. For a weak trace deformation of flat space, the metric source couples linearly to the trace of the stress tensor. The quadratic variation of the matter effective action with respect to this same source is therefore governed by the connected two-point function \(\langle\Theta\Theta\rangle\). Its long-wavelength expansion contains a two-derivative term, which is matched to the corresponding quadratic expansion of the Einstein-Hilbert action. This gives a representation of the induced Newton constant in terms of the scalar trace response, up to contact terms and subtraction ambiguities. In Zee's Yang-Mills application, the trace channel is fixed by the operator-trace anomaly. The massless classical theory has a traceless stress tensor, but the running coupling generates \(\Theta=(\beta(g)/2g)F^a_{\mu\nu}F^{a\mu\nu}\) at the quantum level. This anomalous trace is then inserted into the Adler-Zee response formula, relating the induced Einstein term to the low-energy scalar response of the \(F^2\) channel.\par
The construction used in this section differs from the Adler-Zee setup in both its aim and its organization. We also use the seed CFT path integral via stress-tensor correlation functions or, in the trace channel, via the Weyl-anomaly contact term. However, these data are used to evaluate the first-order response to a non-local stress-tensor deformation, rather than to construct a complete matter effective action whose derivative expansion directly defines an induced Newton constant. A useful feature of the present formulation is that the non-local kernel is part of the deformation data. Once the Weyl-anomaly contact structure is fixed, the local terms obtained after taking the coincidence limit depend on this kernel. For example, choosing the kernel to be the Green function of a Laplace-type operator, possibly with a mass scale or curvature-dependent potential terms, gives a definite derivative expansion. Thus, the anomaly supplies the universal trace-channel input, whereas the kernel specifies how this input is mapped into the first-order gravitational functional.
\subsection{Trace-trace correlator from the Weyl anomaly}
We begin with the standard form of the Weyl anomaly in even dimensions. Let \(\mathcal{W}_{\text{CFT}}[g]\) be the renormalized effective action of a CFT. Under an infinitesimal Weyl transformation \(\delta_\sigma g_{\mu\nu}=2\sigma g_{\mu\nu}\), the variation of the effective action can be written as
\begin{align}
    \delta_\sigma \mathcal{W}_{\text{CFT}}[g] =\int \text{d}^d x\,\sqrt g\,\sigma\,\langle{\Theta}\rangle_{\text{CFT}}= \int \text{d}^d x\,\sqrt g\,\sigma\,{\cal A}_d[g] .
\end{align}
On a smooth manifold without boundary, local Weyl anomalies occur only in
even spacetime dimensions.  In \(d=2n\), the local metric anomaly can be
organized into type-A, type-B, and trivial terms according to the
cohomological and geometric classification of
\cite{Bonora:1983ff,Bonora:1985cq,Deser:1993yx}.  We write
\begin{align}
{\cal A}_{2n}
=
\frac{1}{(4\pi)^n}
\left[
(-1)^{n+1}a\,E_{2n}
+\sum_I c_I\,I_I^{(2n)}
+\sum_J b_J\,D_J^{(2n)}
\right].
\end{align}
Here \(E_{2n}\) is the Euler density,
\begin{align}
E_{2n}
=
\frac{1}{2^n}
\delta^{\mu_1\cdots\mu_{2n}}_{\nu_1\cdots\nu_{2n}}\,
R^{\nu_1\nu_2}{}_{\mu_1\mu_2}
\cdots
R^{\nu_{2n-1}\nu_{2n}}{}_{\mu_{2n-1}\mu_{2n}} ,
\end{align}
with the generalized Kronecker delta totally antisymmetric in both sets of indices.  This term gives the type-A anomaly: its integral is proportional to the Euler characteristic on a closed manifold, but the local density \(\sqrt g\,E_{2n}\) is not pointwise Weyl invariant; rather, its Weyl variation is a total derivative \cite{Deser:1993yx,Duff:1993wm}.  The
scalars \(I_I^{(2n)}\) denote independent local conformal invariants of
dimension \(2n\), equivalently Weyl invariants of weight \(-2n\), so that
\(\sqrt g\,I_I^{(2n)}\) is Weyl invariant in the physical dimension.  These
terms are the type-B anomalies \cite{Deser:1993yx}.  Finally,
\(D_J^{(2n)}\) denote trivial anomalies, namely local terms generated by the
Weyl variation of finite local counterterms; in a covariant curvature basis
they may be chosen as total derivatives, \(D_J^{(2n)}=\nabla_\mu J_J^\mu\)
\cite{Bonora:1985cq,Henningson:1998gx,Duff:1993wm}.  With the normalization
of the basis fixed, the coefficients \(a\) and \(c_I\) multiply
non-trivial Weyl cohomology classes and are scheme independent, whereas the coefficients \(b_J\) depend on the choice of finite local counterterms.
\par
We now specialize the general discussion to four dimensions. In the standard curvature basis, the renormalized Weyl anomaly is written as
\begin{align}\label{four dimensional Weyl anomaly}
{\cal A}_4 = \frac{1}{(4\pi)^2} \left(-a\,E
_4+ c\,W_{\mu\nu\rho\sigma}W^{\mu\nu\rho\sigma}  +\tilde b\,\Box R \right) . 
\end{align}
Here, the four-dimensional Euler density takes the form
\begin{align} 
E_4 = R_{\mu\nu\rho\sigma}R^{\mu\nu\rho\sigma} -4R_{\mu\nu}R^{\mu\nu} +R^2.
\end{align}
The second term is the local Weyl invariant built from two curvatures,
\begin{align}
W_{\mu\nu\rho\sigma}W^{\mu\nu\rho\sigma} = R_{\mu\nu\rho\sigma}R^{\mu\nu\rho\sigma} -2R_{\mu\nu}R^{\mu\nu} +\frac{1}{3} R^2.
\end{align}
The constants \(a\) and \(c\), multiplying the type-A and type-B terms respectively, are the scheme-independent central charges associated with the stress-tensor sector of the four-dimensional CFT. The term proportional to \(\tilde b\) is the four-dimensional trivial anomaly. Its coefficient is scheme dependent, since a finite local \(R^2\) counterterm shifts the coefficient of \(\Box R\) in the anomaly. Since it only affects scheme-dependent local contact terms, we will set \(\tilde b=0\) when extracting the universal part of the trace-channel response.\par
We now compute the contact terms in the \(\Theta\Theta\) correlator. As in \cite{Hartman:2023qdn, Hartman:2023ccw}, the trace-trace contact term is obtained by varying the trace one-point function with respect to a local Weyl rescaling of the metric,
\begin{align}
\langle \Theta(x)\Theta(y)\rangle_{\text{CFT}} &= \frac{4}{\sqrt{g(x)}\sqrt{g(y)}} g_{\mu\nu}(x) \frac{\delta}{\delta g_{\mu\nu}(x)} \left[ g_{\rho\sigma}(y) \frac{\delta {\cal W}_{\rm CFT}[g]}{\delta g_{\rho\sigma}(y)} \right] \notag\\ &= \frac{1}{\sqrt{g(x)}\sqrt{g(y)}}\frac{\delta}{\delta\sigma(x)} \left[ \sqrt{g(y)}\,\langle\Theta(y)\rangle_{\text{CFT}} \right] . 
\end{align}
The trace one-point function in a CFT is given by the Weyl anomaly, $\langle{\Theta}\rangle_{\text{CFT}}=\mathcal{A}_{d}$.  Taking $d=4$ as a concrete example, the trace–trace two-point function is obtained by functionally differentiating (\ref{four dimensional Weyl anomaly}) with respect to the Weyl factor,
\begin{align}\label{tracetrace two-point function}
    \langle \Theta(x)\Theta(y)\rangle_{\text{CFT}} &=-\frac{8a}{(4\pi)^2}G^{\mu\nu}(x)\nabla^{(x)}_{\mu}\nabla^{(x)}_{\nu}\tilde\delta(x,y)+\tilde b\,\text{-terms},
\end{align}
where $G^{\mu\nu}=R^{\mu\nu}-\frac{1}{2}Rg^{\mu\nu}$ is the Einstein tensor, and  \(\tilde\delta(x,y)\) is the covariant delta function.
\subsection{Trace-trace deformation and scheme-independent gravitational action}
Having determined the contact structure of the \(\langle\Theta\Theta\rangle\)
correlator from the Weyl anomaly, we now turn to its role in the stress tensor deformed
theory.  Following the strategy developed in the previous sections, we consider
a non-local trace-trace deformation of the seed CFT, and evaluate the
first-order correction to the effective action. A related non-local trace-trace deformation was discussed in our previous work \cite{Xie:2026kek}. There, the non-local insertion was chosen in a specific form adapted to the explicit structure of the \(\langle\Theta\Theta\rangle\) correlator, and the induced gravitational action was organized as an expansion in the UV cutoff. In the present section, we revisit this construction in a more systematic form. The non-local kernel is now treated as part of the deformation data and may be chosen as the inverse of a general minimal operator. Moreover, instead of keeping the cutoff expansion itself, we follow the standard renormalization procedure and extract the finite scheme-independent gravitational action.\par
We deform the seed CFT by a non-local bilinear built from the trace of the stress tensor. The corresponding flow of the action is taken to be 
\begin{align}
\partial_{\lambda} S^{(\lambda)} = \int_{{\cal M}\times{\cal M}} \text{d}\mu(x)\,\text{d}\mu(y)\, \Theta(x)H(x,y)\Theta(y) ,
\end{align} 
where \(\text{d}\mu(x)=\text{d}^d x\sqrt{g(x)}\). The kernel \(G(x,y)\) is specified as part of the deformation data. The first-order correction of the deformed effective action is given by
\begin{align}
    \mathcal{W}^{(1)}=\lambda\int \text{d}\mu(x)\text{d}\mu(y)\,H(x,y)\langle{\Theta(x)\Theta(x)}\rangle^{(0)}.
\end{align}
In four-dimensional spacetime, for example, the trace-trace two-point function (\ref{tracetrace two-point function}) is substituted into the above equation, and a straightforward calculation yields
\begin{align}\label{4dW1SI}
    \mathcal{W}^{(1)}_{d=4}=-\frac{8a\lambda}{(4\pi)^2}\int \text{d}\mu(y)\lim_{x\to y}\Big[G^{\mu\nu}(x)\nabla_{\mu}^{(x)}\nabla_{\nu}^{(x)}H(x,y)\Big]+\tilde b\,\text{-terms}.
\end{align}
In order to express the first-order correction as a combination of geometric invariants, it is necessary to specify the explicit form of the non-local kernel $H(x,y)$. As a simple representative choice, we take $H(x,y)$ to be the Green's function of a second-order Laplace-type operator, satisfying
\begin{align}\label{GEq 105}
\Big(-\Box_x+N(x)\Big)H(x,y) = \tilde\delta(x,y). 
\end{align} 
Here \(N(x)\) is a local potential term. The restriction to a second-order kernel is only for notational simplicity; the same analysis can be extended to higher-order minimal operators. Through the Mellin transform, $H(x,y)$ can be expressed in terms of the heat kernel of the Laplace-type operator, which admits a small-$\tau$ expansion,
\begin{align}
    H(x,y)=\int_{0}^{\infty}\text{d}\tau\,\frac{\Delta^{1/2}(x,y)}{(4\pi\tau)^{d/2}}e^{-\sigma(x,y)/2\tau}\sum_{n=0}^{\infty}a_{n}(x,y)\tau^{n}+H_{\text{global}}(x,y).
\end{align}
In this paper, we focus on the UV behavior of the heat kernel. As in the previous section, an upper limit $\tau = L^2$ is introduced to exclude IR divergences, and dimensional regularization $d = 4 - \varepsilon$ is employed. The regularized Green's function takes the form
\begin{align}\label{HUVreg}
    H^{(\varepsilon)}_{\text{UV,reg}}(x,y)&=\frac{\Delta^{1/2}(x,y)}{(4\pi)^2}\sum_{n=0}^{\infty}a_{n}(x,y)\mu^{\varepsilon}\Big[\Big(\frac{\sigma(x,y)}{2}\Big)^{n-1+\varepsilon/2}\Gamma(1-n-\varepsilon/2)\notag\\
    &\quad-\sum_{m=0}^{\infty}\frac{(-1)^mL^{2(n-1)-2m+\varepsilon}}{m!(m+1-n-\varepsilon/2)}\Big(\frac{\sigma(x,y)}{2}\Big)^m\Big].
\end{align}
Next, we compute the first-order correction to the effective action in (\ref{4dW1SI}). The first term in (\ref{HUVreg}) corresponds to the non-analytic part and requires careful treatment in the coincidence limit. Writing \(z=\sigma/2\), its \(n\)-th term is proportional to \(z^{\,n-1+\varepsilon/2}\Gamma(1-n-\varepsilon/2)\). After the action of two covariant derivatives in (\ref{4dW1SI}), the leading singular behavior is of order \(z^{\,n-2+\varepsilon/2}\). For \(\operatorname{Re}\varepsilon>0\), the ordinary coincidence limit of the non-analytic term vanishes for \(n\geq 3\). The \(n=2\) term is proportional to \(z^{\varepsilon/2}\Gamma(-1-\varepsilon/2)\). With \(\operatorname{Re}\varepsilon>0\) fixed, \(z^{\varepsilon/2}\to0\) as \(x\to y\), so this term has no finite diagonal value. If one expands in \(\varepsilon\) first, however, the same factor generates a pole and a \(\log z\) singularity. This is a short-distance singularity rather than a regular coincidence
coefficient, defined only within the full regulated kernel. The \(n=1\) term gives a locally integrable short-distance singularity of order \(z^{-1+\varepsilon/2}\), which does not carry an independent contact residue on the diagonal. The \(n=0\) term is singular already at fixed \(\varepsilon\): after two covariant derivatives it behaves as \(z^{-2+\varepsilon/2}\) near the diagonal. It therefore does not define a regular coincidence coefficient. Rather, it belongs to the singular Green's function parametrix, whose role is to reproduce the diagonal source of the defining operator. The local UV density should consequently be extracted from the complete regulated kernel, with the \(\varepsilon/2\) shift kept until the final Laurent expansion. The second term in (\ref{HUVreg}) is handled in the coincidence limit by ordinary methods. Putting everything together, we obtain the regularized first-order correction to the effective action as
\begin{align}\label{W1UVreg}
    \mathcal{W}^{(1)}_{\text{UV,reg}}&=-\frac{8a\lambda}{(4\pi)^2}\mu^{\varepsilon}\sum_{n=0}^{\infty}\int \text{d}\mu(y)\bigg\{\frac{L^{2n-2+\varepsilon}}{n-1+\varepsilon/2}\Big(G^{\mu\nu}[\nabla_{\mu}\nabla_{\nu}a_n](y)+\frac{1}{6}R_{\mu\nu}G^{\mu\nu}[a_n](y)\Big)\notag\\
    &\quad+\frac{2-\varepsilon}{4}\frac{L^{2n-4+\varepsilon}}{n-2+\varepsilon/2}R[a_n](y)\bigg\}+\cdots.
\end{align}
Equivalently, as in the previous sections, we may take the coincidence limit before carrying out the proper-time integral. For fixed \(\tau>0\), the heat-kernel integrand is regular near the diagonal, so the limit can be taken term by term. The subsequent proper-time integrations then reproduce exactly the regularized expression obtained above.\par
Next, (\ref{W1UVreg}) is expanded in powers of \(\varepsilon\) and the logarithmic divergence is extracted. After renormalization, a scheme-independent finite term is obtained from the part that does not depend on the arbitrary scale \(L\),
\begin{align}
    \mathcal{W}^{(1)}_{\text{SI}}&=-\frac{16a\lambda\ln(\mu/\mu_0)}{(4\pi)^2}\int \text{d}\mu(y)\Big(G^{\mu\nu}[\nabla_{\mu}\nabla_{\nu}a_1](y)+\frac{1}{6}R_{\mu\nu}G^{\mu\nu}[a_1](y)+\frac{1}{2}R[a_2](y)\Big).
\end{align}
The coincidence limit of the heat kernel coefficients is determined by the data specifying the non-local kernel of the deformation. For a concrete example, we choose $N(x)=m^{2}$ in (\ref{GEq 105}) and express the first-order correction to the effective action as a linear combination of geometric invariants,
\begin{align}
    \mathcal{W}^{(1)}_{\text{SI}}&=-\frac{8a\lambda\ln(\mu/\mu_0)}{(4\pi)^2}\int \text{d}\mu(y)\Big(\frac{1}{2}m^4R-\frac{1}{3}m^2R_{\mu\nu}R^{\mu\nu}+\frac{1}{60}R\Box R+\frac{1}{30}R_{\mu\nu}\Box R^{\mu\nu}\notag\\
    &\quad-\frac{1}{120}R^3+\frac{1}{60}RR_{\mu\nu}R^{\mu\nu}+\frac{1}{15}R_{\mu\nu}R_{\rho\sigma}R^{\rho\nu\sigma\mu}\Big).
\end{align}

\section{Conclusion and outlook}\label{sec:conclusion}
In this work, we have studied the effective gravitational action generated, to first order in the deformation parameter, by non-local $T\bar T$-like stress-tensor deformations. Starting from the quantum stress-tensor data of the seed theory, we developed a heat-kernel framework for extracting local geometric terms induced by the deformation. For general free field theories, this provides a unified way to compute the contact contribution of the stress-tensor two-point function and to rewrite its contraction with the non-local kernel as a local gravitational density. The examples of free fermions, massive Maxwell theory, and second-order Yang-Mills theory illustrate how the construction extends beyond the scalar case considered previously, while also showing that the resulting coefficients are, in general, sensitive to the microscopic content of the seed theory. The main theory-independent results were obtained for conformal seed theories. In this case, conformal symmetry fixes the leading non-contact part of the stress-tensor two-point function up to the central charge $C_T$, allowing us to extract universal contributions to the induced gravitational action without specifying a particular Lagrangian model. After regularization and renormalization, these contributions give finite, scheme-independent geometric terms organized by a finite set of curvature invariants. We also analyzed the trace sector, where the separated-point correlator vanishes in a CFT, and the remaining contact contribution is fixed by the Weyl anomaly. For general choices of the non-local kernel within the class considered, this anomaly-controlled sector yields additional finite contributions to the effective action. Taken together, our results show that non-local stress-tensor deformations provide a quantum effective-action realization of the induced-gravity idea: universal CFT data determine calculable sectors of the emergent geometric response, while model-dependent correlator data and scheme-dependent counterterms remain clearly separated.\par
Several directions appear particularly important for extending the present analysis. A first priority is to go beyond the first-order deformation parameter. While the present work identifies renormalized local contributions to the induced gravitational action at leading order, it remains to be understood how these terms are reorganized at higher orders, and to what extent they continue to admit a controlled geometric interpretation once separated-point data and genuinely non-local effects are iterated. A second direction is to enlarge the universal conformal sector beyond the contributions controlled by $C_T$ and the Weyl anomaly. In particular, it would be important to determine how subleading terms in the local $TT$ OPE, renormalized higher-point stress-tensor correlators, mixed correlators, and possible background- or state-dependent one-point functions enter the induced action, and which of the resulting geometric structures remain genuinely universal. Closely related is the problem of consistency conditions on the non-local kernel itself. Since the deformation is intrinsically irrelevant and, in dimensions higher than two, generically non-local, a sharper understanding of the constraints imposed by unitarity, causality, positivity, and analyticity should help distinguish admissible deformation data from merely formal choices. Finally, it would be valuable to extend the construction to more general operators and geometric settings, including non-minimal kernels, manifolds with boundaries or defects, and observables beyond the vacuum effective action, such as correlation functions, thermal partition functions, and entanglement measures on curved backgrounds. Progress on these questions should provide a more complete test of the induced-gravity interpretation developed here and clarify the extent to which stress-tensor deformations can furnish a systematic quantum route to emergent geometry.
\section*{Acknowledgments}
We thank Shan-Wen Jiang, Yunfei Xie, Long Zhao, and Yu-Xuan Zhang for helpful discussions. This work was supported by the National Natural Science Foundation of China (Grants No. 12475053, No. 12475056, No. 12588101, No. 12235016, and No. 12247101), Gansu Province’s Top Leading Talent Support Plan, the Fundamental Research Funds for the Central Universities (Grant No. lzujbky-2025-jdzx07), the Natural Science Foundation of Gansu Province (No. 22JR5RA389 and No. 25JRRA799), and the 111 Project (Grant No. B20063).
\bibliographystyle{JHEP}
\bibliography{scalartoLiouville}

\end{document}